\documentclass{article}

\usepackage{arxiv}

\usepackage[utf8]{inputenc} 
\usepackage[T1]{fontenc}    
\usepackage{hyperref}       
\usepackage{url}            
\usepackage{booktabs}       
\usepackage{amsmath}
\usepackage{amsfonts}       
\usepackage{nicefrac}       
\usepackage{microtype}      
\usepackage{graphicx}
\usepackage{natbib}
\usepackage{doi}

\usepackage{xspace}
\newcommand{\ace}{ACE2.5-ERA5\xspace}

\title{Stochastic climate emulation at reduced cost with \ace}

\author{
  Jeremy McGibbon\thanks{Corresponding author: \texttt{jeremym@allenai.org}},
  Troy Arcomano,
  Spencer K.\ Clark,
  James P.\ C.\ Duncan, \\
  \textbf{Brian Henn,
  Oliver Watt-Meyer}\\[0.3em]
  \mdseries Allen Institute for Artificial Intelligence, Seattle, WA, USA
}

\renewcommand{\headeright}{}
\renewcommand{\undertitle}{}
\renewcommand{\shorttitle}{Stochastic climate emulation at reduced cost with \ace}

\hypersetup{
pdftitle={Stochastic climate emulation at reduced cost with ACE2.5-ERA5},
pdfsubject={physics.ao-ph, cs.LG},
pdfauthor={Jeremy McGibbon, Troy Arcomano, Spencer Clark, James Duncan, Brian Henn, Elynn Wu, Oliver Watt-Meyer, Christopher S. Bretherton},
pdfkeywords={climate emulation, stochastic weather prediction, machine learning, CRPS, ensemble calibration},
}

\begin{document}
\maketitle

\begin{abstract}
We introduce \ace (Ai2 Climate Emulator of ERA5, version 2.5), a stochastic autoregressive weather-climate AI model that produces calibrated weather ensembles with realistic small-scale structure on timescales from days to centuries. Its deterministic predecessor, ACE2-ERA5, retains 36\% of its ERA5 target's precipitation spectral power at total wavenumbers 100--179; \ace retains 86\%. In weather forecasts its ensemble spread matches its error to within 4\% from day 3. Time-mean climate is on par with or better than ACE2-ERA5, the precipitation response to ENSO is improved, and the model is stable over 1000 years. Interannual variability of global mean surface temperature is shown not to be calibrated in either \ace or its predecessor. \ace differs from ACE2-ERA5 in three ways. Noise-conditioned layer normalization, trained with a combined CRPS and spectral energy score loss in place of mean squared error, makes it stochastic. Spectral bottlenecking with grouped convolutions cuts the Spherical Fourier Neural Operator's spectral filter parameters by more than two orders of magnitude and leaves a model a tenth the size of ACE2-ERA5. Its time step is daily rather than 6-hourly. \ace has 3.2$\times$ the inference throughput of ACE2-ERA5. A 6-hourly variant has weather prediction skill similar to ACE2-ERA5. A 4-degree variant reaches comparable climate skill at an order of magnitude lower training cost, enabling rapid iteration during development.
We release 1-degree daily, 4-degree daily and 1-degree 6-hourly checkpoints.
\end{abstract}


\section{Introduction}

Machine learning models have become competitive with physics-based numerical weather prediction \citep{lam23, price25, bonev25}, and their low computational cost makes them attractive for climate applications that require long integrations or large ensembles. ACE2-ERA5 \citep{wattmeyer25} showed that a machine learning emulator reproduces atmospheric climate statistics from days to decades when forced with observed sea surface temperature, sea ice and CO$_2$. Later work has applied stochastic versions of ACE2's architecture to emulation of a global storm-resolving model coupled with diffusion-based downscaling \citep{perkins26}, to emulation of a fully coupled Earth system model \citep{wu26}, and to separating the atmospheric response to CO$_2$ and sea surface temperature \citep{clark2026}. Here we introduce \ace, a stochastic version of ACE2 that generates calibrated ensemble predictions with realistic spatial structure on timescales from days to centuries, at substantially lower computational cost than ACE2.
A related model, ACE2.2-ERA5, adapts these developments to the AIMIP protocol \citep{aimip26}.

ACE2-ERA5 is deterministic: each forward step maps an input state to one output state. Deterministic models trained with mean squared error regress toward the conditional mean, suppressing grid-scale variability that matters for local climate assessment. ACE2-ERA5 retains only 36\% of ERA5's spectral power of 6-hourly precipitation at total wavenumbers 100--179, making its output difficult to use in statistical downscaling. Local precipitation extremes are also underestimated. For climate projection, where the interest is often in changes to regional precipitation intensity or spatial patterns of temperature variability, this loss of small-scale structure is a fundamental limitation.

Several stochastic weather models have addressed this problem by replacing MSE with the continuous ranked probability score (CRPS), a proper scoring rule for probabilistic forecasts. AIFS-CRPS \citep{lang24}, GenCast \citep{price25}, and FGN \citep{alet25} each produce calibrated ensemble forecasts with realistic variability. However, CRPS evaluated pointwise does not constrain the spatial structure of the ensemble: models trained on CRPS alone tend to produce excess spectral power at small scales \citep{alet25, swift}, requiring either architectural filtering \citep{lang24} or additional spectral loss terms \citep{bonev25, kochkov24} to suppress the noise. None of these models have been evaluated as climate emulators on multi-decadal timescales, where compounding errors are particularly important. Notably, ArchesWeatherGen \citep{couairon26} is a stochastically autoregressive model evaluated on climate timescales \citep{singh26} which uses a denoising loss function rather than CRPS.

\ace makes three substantial changes to ACE2. First, we introduce noise-conditioned layer normalization and train with a combined CRPS and spectral energy score loss, where the spectral term penalizes discrepancies in both spectral power and phase at every wavenumber, suppressing small-scale noise (Section~\ref{sec:loss}). Second, we reduce the Spherical Fourier Neural Operator's spectral filter parameters by more than two orders of magnitude through spectral bottlenecking and grouped convolutions, which makes the model a tenth the size of ACE2-ERA5, faster to train and faster to run (Section~\ref{sec:bottlenecking}). Third, we move from a 6-hourly to a daily forward step, as done in AIFS-SUBS \citep{aifssubs26}. For applications that do not require an explicit diurnal cycle this is the larger part of \ace's gain in inference throughput, and it would also reduce the cost of coupled atmosphere-ocean training with ocean models that have daily or longer time steps \citep{wu26, duncan26}.

\ace tracks the observed warming over 1940--2020 with a global-mean 2\,m temperature error of 0.05\,K, improves the precipitation response to ENSO, and runs stably for 1000 years under repeating climatological forcing. Its daily precipitation retains realistic spectral power (a significant deficit in ACE2-ERA5) and reproduces the frequency of extreme precipitation events (Section~\ref{sec:precip_structure}). Climate skill is consistent across the four random seeds we trained, so no seed selection is needed (Appendix~\ref{sec:seed-variability}). We release the 1-degree daily, 4-degree daily, and 1-degree 6-hourly checkpoints for use in downstream applications.

\section{Evaluation}
\label{sec:evaluation}

We evaluate three configurations of the model against the published deterministic ACE2-ERA5 \citep{wattmeyer25}, which has a 1-degree grid and a 6-hourly time step. \ace is the 1-degree daily configuration, \ace-4deg the 4-degree daily configuration, and \ace-6h the 1-degree 6-hourly configuration (Table~\ref{tab:runs}); figure legends label them ``1deg daily'', ``4deg daily'' and ``1deg 6h''. The three configurations use the same architecture and loss, and they are stochastic, so each run is one draw from an ensemble. Section~\ref{sec:architecture} describes the architecture and training recipe, and Appendix~\ref{sec:ablation-figures} the ablations that support its design choices.

\begin{table}[ht]
\centering
\caption{Models evaluated in this section. Members are ensemble members per weather initial condition; the 10-year, AMIP and 1000-year runs use one member per initial condition. Ablation variants are listed in Table~\ref{tab:ablations}.}
\label{tab:runs}
\small
\begin{tabular}{lllllrr}
\toprule
Model & Figure label & Resolution & Time step & Loss & Seeds & Members \\
\midrule
\ace & 1deg daily & 1$^\circ$ & 24\,h & CRPS + spectral energy score & 4 & 9 \\
\ace-4deg & 4deg daily & 4$^\circ$ & 24\,h & CRPS + spectral energy score & 4 & 9 \\
\ace-6h & 1deg 6h & 1$^\circ$ & 6\,h & CRPS + spectral energy score & 1 & 9 \\
ACE2-ERA5\textsuperscript{$\dagger$} & ACE2-ERA5 & 1$^\circ$ & 6\,h & MSE & 1 & 1 \\
\bottomrule
\end{tabular}
\par\smallskip
{\footnotesize \textsuperscript{$\dagger$}\citet{wattmeyer25}}
\end{table}

Every model is evaluated with the protocol of \citet{wattmeyer25}, which has four types of run (Table~\ref{tab:protocol}). The 10-year runs cover the last decade of the years withheld from training, and the weather forecasts cover the last withheld year (Section~\ref{sec:training}).

\begin{table}[ht]
\centering
\caption{Evaluation runs. Observed forcing is ERA5 sea surface temperature and sea ice with historical CO$_2$. AMIP runs end in 2025 for \ace models and in 2020 for ACE2-ERA5. Weather forecasts hold sea surface temperature, sea ice and CO$_2$ at their initial values. In weather runs, stochastic models produce 9 ensemble members per initial condition, while they produce 1 ensemble member per initial condition for other run types.}
\label{tab:protocol}
\small
\begin{tabular}{lllp{0.3\textwidth}}
\toprule
Run & Initial conditions & Length and forcing & Used for \\
\midrule
10-year & 2001-01-01, 02, 03 at 00Z & to 2010-12-31, observed & time-mean climate, precipitation, ENSO, cyclones \\
AMIP & 1940-01-03, 04, 05 & to 2025-12-31, observed & climate evolution, CO$_2$ sensitivity \\
Weather & 52 weekly, 00Z, in 2020 & 15 days, held at initial values & calibration, forecast skill \\
1000-year & one & repeating 1991--2020 climatology & stability \\
\bottomrule
\end{tabular}
\end{table}

Three conventions apply to every figure in this section.
\begin{itemize}
\item \emph{Each model is scored against the ERA5 dataset it was trained on}, at its own resolution and time step: daily models on daily means (their daily-mean diagnostic outputs) and 6-hourly models on the time mean of their 6-hourly snapshots.
\item \emph{Figures show one random seed per model}, the first trained, unless stated otherwise. \ace and \ace-4deg were each trained with four seeds, which give comparable results (Appendix~\ref{sec:seed-variability}). Bar charts show the mean over three initial conditions (not seeds) with a 95\% interval; in weather forecast runs, stochastic models produce nine ensemble members per initial condition, and the plotted metric is the ensemble mean (Table~\ref{tab:protocol}).
\item \emph{Hatched bar segments are errors truncated to the 4-degree grid.} A coarser grid has less variance to get wrong, so \ace-4deg's RMSE is not directly comparable with the 1-degree models except in global means. The RMSE bar charts therefore also show each 1-degree model's error after spectral truncation to the 4-degree grid, which is the number to compare with \ace-4deg.
\end{itemize}

\subsection{Time-mean climate}

\begin{figure}
\centering
\includegraphics[width=\textwidth]{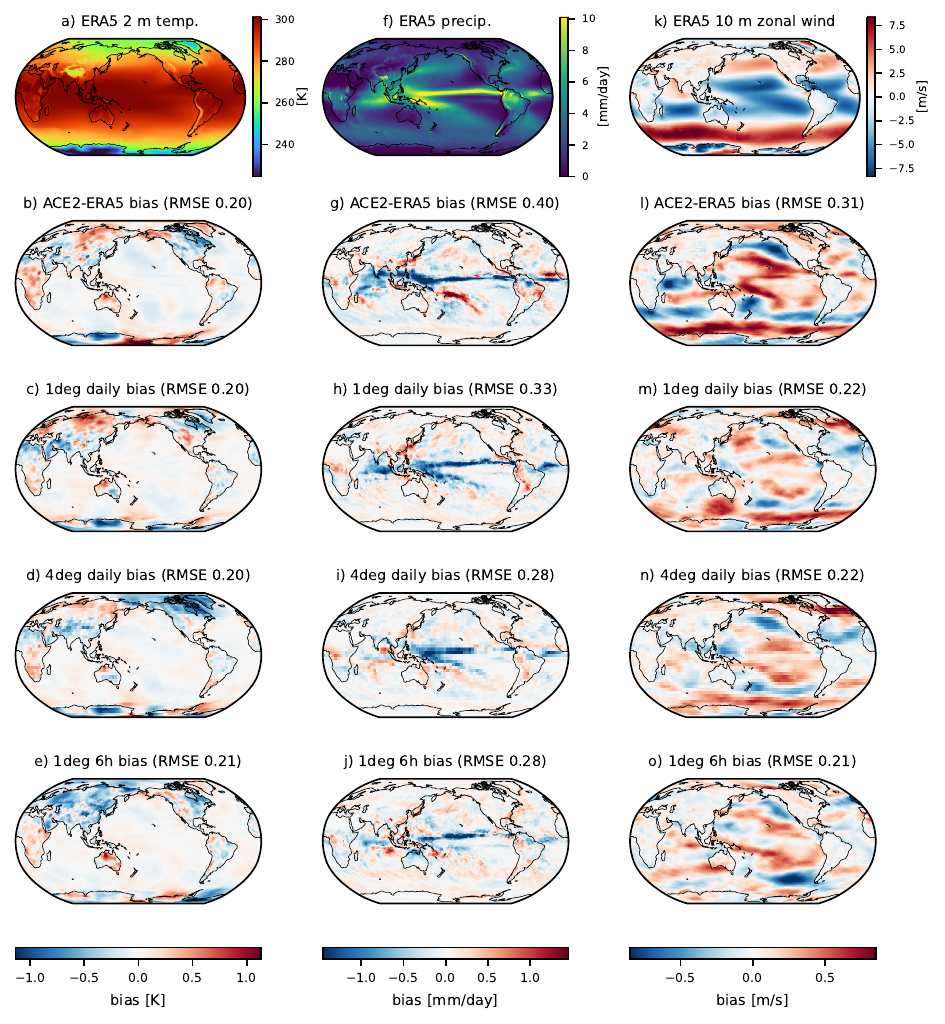}
\caption{Time-mean bias over the 10-year run from 2001-01-01 for 2\,m temperature, surface precipitation rate and 10\,m zonal wind. The top row is the ERA5 time mean; each other row is one model's bias against ERA5 with its area-weighted RMSE in the panel title.}
\label{fig:bias_maps}
\end{figure}

\begin{figure}
\centering
\includegraphics[width=0.85\textwidth]{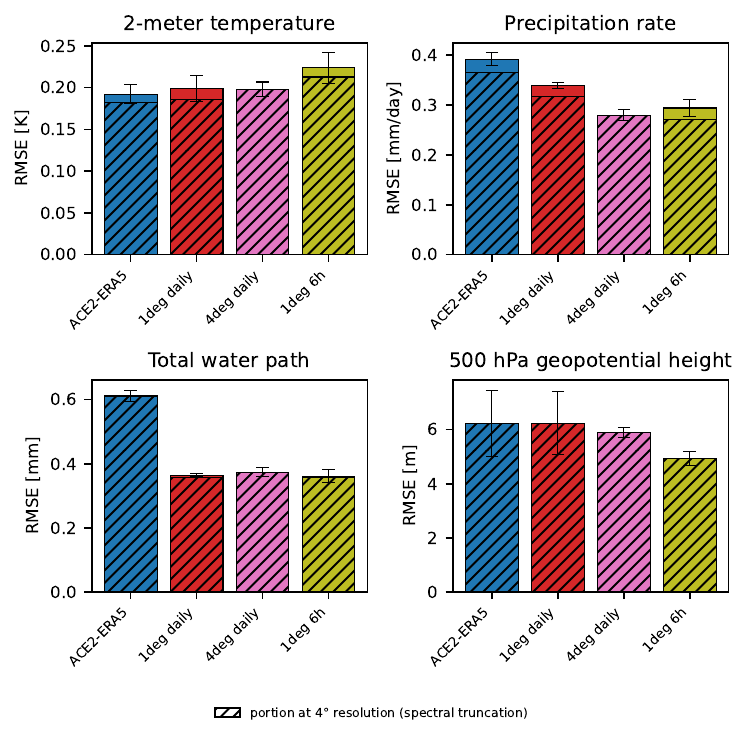}
\caption{Area-weighted RMSE of the 10-year time mean against ERA5 for \ace, \ace-4deg, \ace-6h and ACE2-ERA5, averaged over initial conditions (error bars: 95\% interval). The hatched lower segment of each bar is the RMSE after the bias map is truncated to the 4$^\circ$ grid by a spherical-harmonic round-trip, and the unhatched remainder is the additional error at scales the 4$^\circ$ grid does not resolve; \ace-4deg's bar is entirely hatched.}
\label{fig:time_mean_rmse}
\end{figure}

The changes in \ace improve the time-mean precipitation and near-surface zonal wind without degrading 2m temperature compared with ACE2-ERA5 (Figs.~\ref{fig:bias_maps} and \ref{fig:time_mean_rmse}). \ace reduces the time-mean RMSE of total water path by about 40\% and of precipitation by 13\% relative to ACE2-ERA5, while 2\,m temperature and 500\,hPa height are comparable. \ace-4deg improves further on precipitation, with most of that gap surviving when \ace's bias is truncated to the 4$^\circ$ grid, so it is not an effect of scoring on a coarser grid. The precipitation gain is on the equator and in the South Pacific Convergence Zone; both models share a dry band along the Pacific ITCZ. \ace's 10\,m wind bias is smaller in the storm tracks, and its temperature bias slightly larger over high-latitude land.

\ace-6h improves on the daily model in precipitation (scored as the daily accumulation of the four 6-hourly values), matches it in total water path and achieves the lowest 500\,hPa height RMSE, while its 2\,m temperature is comparable to \ace.

Two further checks from \citet{wattmeyer25} are in Appendix~\ref{sec:si-diagnostics}. The lag correlation of 20--100-day filtered tropical precipitation reproduces the eastward propagation of the Madden--Julian Oscillation, with a weaker signal east of 120$^\circ$E than either ERA5 or ACE2-ERA5 (Figure~\ref{fig:mjo}). \ace and ACE2-ERA5 both reproduce the annual cycle of the stratospheric vortex at 60$^\circ$N and 60$^\circ$S, and neither differs significantly from ERA5 in its 10-year climatological mean (Figure~\ref{fig:strat}).

\subsection{Precipitation structure}
\label{sec:precip_structure}

\begin{figure}
\centering
\includegraphics[width=\textwidth]{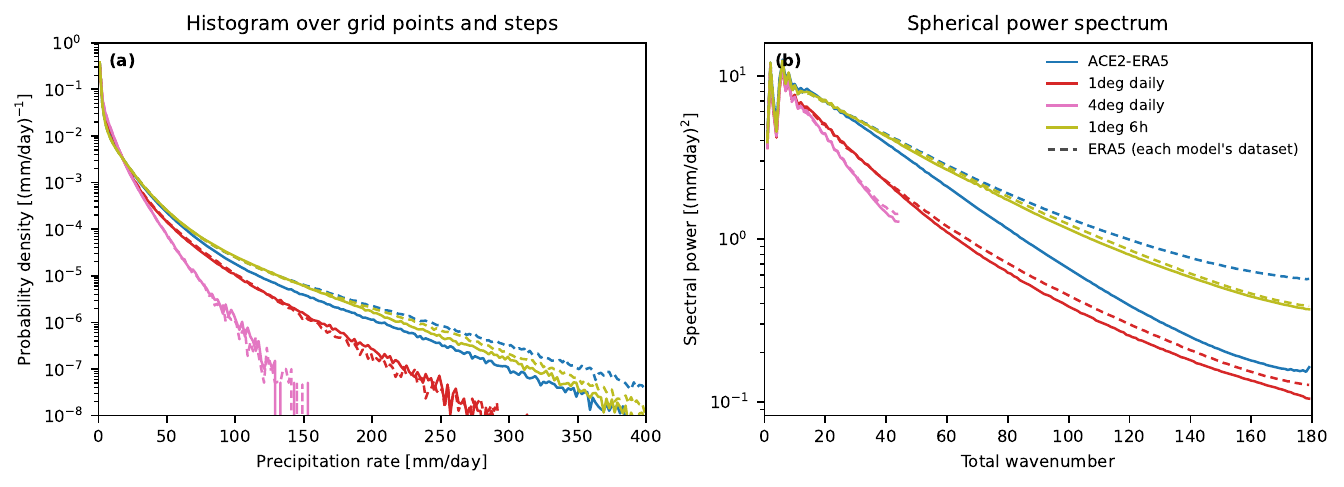}
\caption{Area-weighted histogram (a) and mean spherical power spectrum (b) of surface precipitation rate over a 10-year run, each model at its native output time step (solid) with the ERA5 dataset it was trained on in the same color (dashed). The three 1-degree ERA5 lines differ because the datasets differ at the smallest scales (Appendix~\ref{sec:si-data}): the \ace daily dataset, the \ace 6-hourly dataset, and the ACE2 6-hourly dataset.}
\label{fig:power_spectrum_comparison}
\end{figure}

\begin{figure}
\centering
\includegraphics[width=\textwidth]{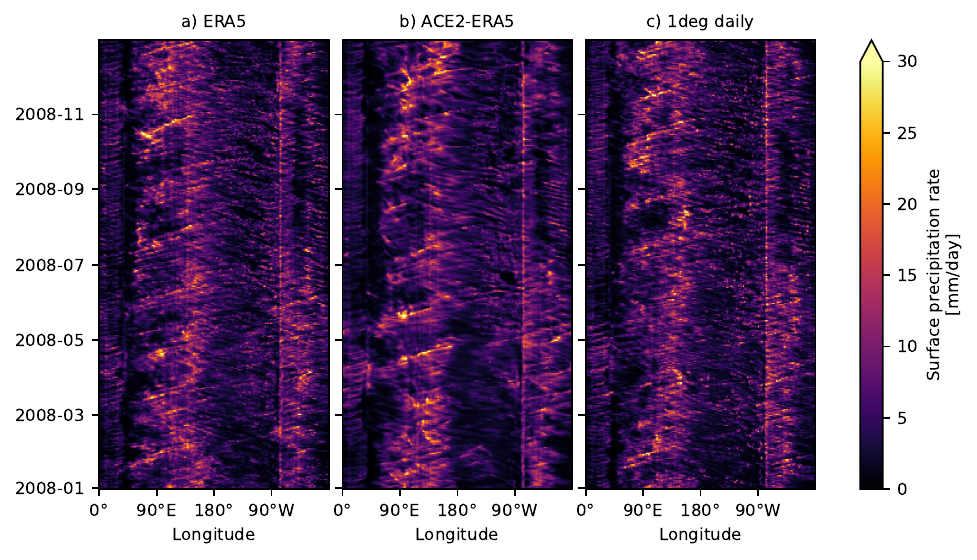}
\caption{Daily precipitation averaged over 10$^\circ$S--10$^\circ$N during 2008 of the 10-year run from 2001-01-01, for ERA5, ACE2-ERA5, and \ace.}
\label{fig:hovmoller}
\end{figure}

A model trained with mean squared error is rewarded for predicting the average of the outcomes it cannot tell apart, which smooths the fields it produces. The smoothing is strongest for precipitation at the grid scale, where the outcome is least predictable. Stochastic training removes that incentive: each ensemble member is a sample, and the loss rewards samples that are distributed like the data.

The power spectrum shows the result (Figure~\ref{fig:power_spectrum_comparison}b). Over total wavenumbers 100--179, the smallest scales the 1-degree grid resolves, ACE2-ERA5 retains 36\% of its ERA5 target's spectral power and \ace retains 86\%. Both models match their targets at the largest scales. ACE2-ERA5's deficit opens near wavenumber 40 and grows toward the grid scale, where the deterministic model's regression to the mean is most severe. \ace-6h retains 94\% of its 6-hourly target's power over the same range, more than the daily model. Six-hourly precipitation has a flatter spectrum than the daily mean, which averages out much of the convective variability, so small scales contribute more to the loss.

The recovered variance is in the right places. \ace reproduces the frequency of days exceeding 100\,mm/day, where ACE2-ERA5 underestimates the frequency of the heaviest 6-hourly rates in its own target (Figure~\ref{fig:power_spectrum_comparison}a). In the tropics, its daily precipitation predominantly propagates eastward in localized structures like ERA5's, where ACE2-ERA5 produces smoother and broader bands (Figure~\ref{fig:hovmoller}).

\subsection{Climate evolution}

\begin{figure}
\centering
\includegraphics[width=\textwidth]{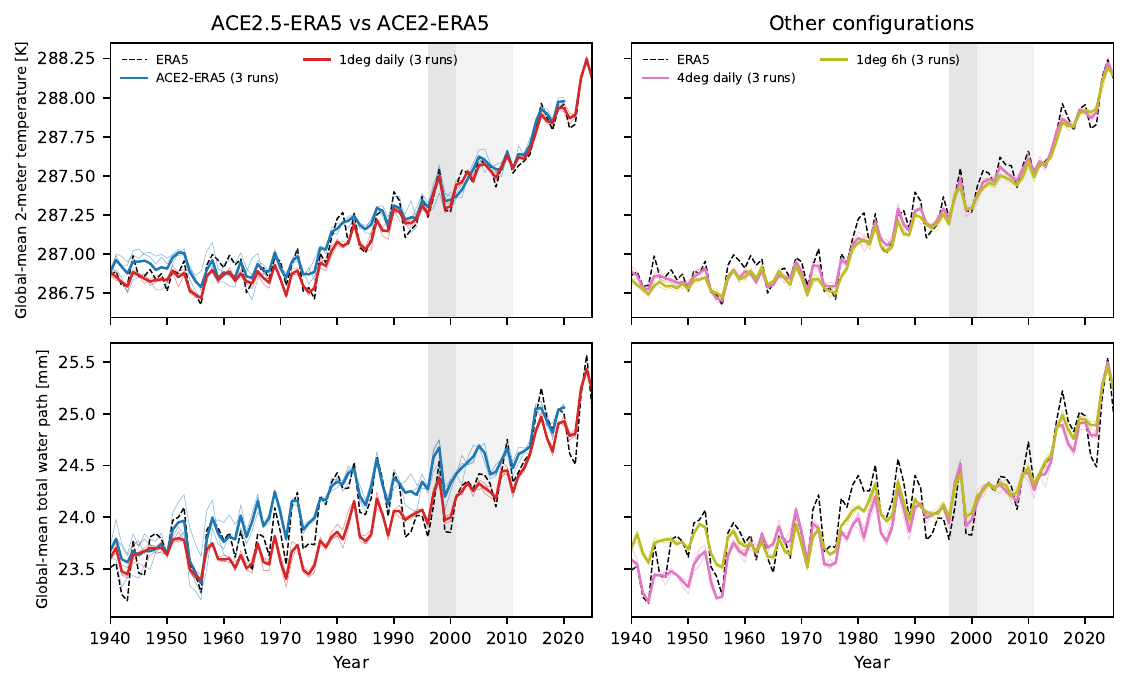}
\caption{Annual global-mean 2\,m temperature (top) and total water path (bottom) over the AMIP runs initialized in January 1940, compared with ERA5 (dashed), one line per run. \ace models run through 2025; ACE2-ERA5 through 2020. Shading marks the years withheld from training; the lighter band (2001--2010) is also the evaluation window of the 10-year runs.}
\label{fig:climate_evolution}
\end{figure}

\begin{figure}
\centering
\includegraphics[width=0.45\textwidth]{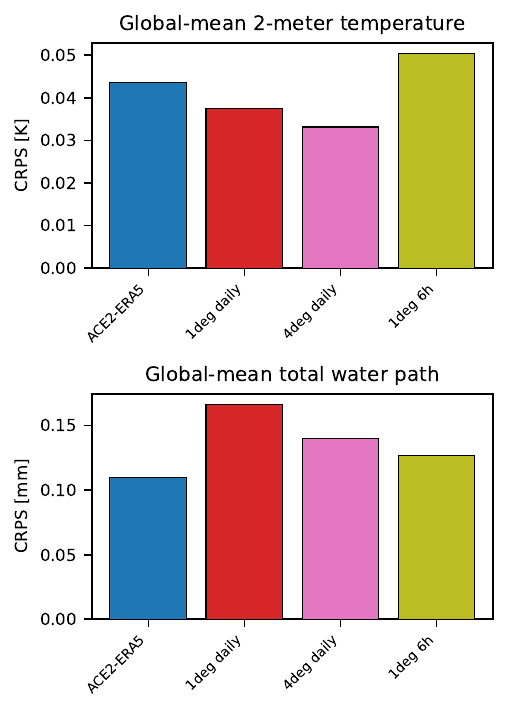}
\caption{Fair CRPS of the annual global-mean 2\,m temperature and total water path over the full AMIP runs, treating each model's initial conditions as an ensemble and scoring against ERA5. A global mean is the same at any resolution, and ACE2-ERA5 produces an initial condition ensemble, so all configurations shown here are directly comparable.}
\label{fig:climate_evolution_crps}
\end{figure}

The AMIP runs test whether a model follows the observed atmospheric warming under observed SST forcing. \ace follows the 2\,m temperature record more closely than ACE2-ERA5 (Figure~\ref{fig:climate_evolution}), and \ace-4deg tracks it more closely still. \ace's total water path is less faithful: it runs slightly dry through most of the 1970s and 1980s, while ACE2-ERA5 runs slightly moist during 1992--2009. \ace-4deg repeats \ace's offsets.

Figure~\ref{fig:climate_evolution_crps} scores the same series with fair CRPS, treating a model's initial conditions as an ensemble forecast of each year's global mean. The temperature score favours \ace and \ace-4deg over ACE2-ERA5; the total water path score favours ACE2-ERA5, reflecting \ace's dry offset in the 1970s and 1980s.

\begin{figure}
\centering
\includegraphics[width=\textwidth]{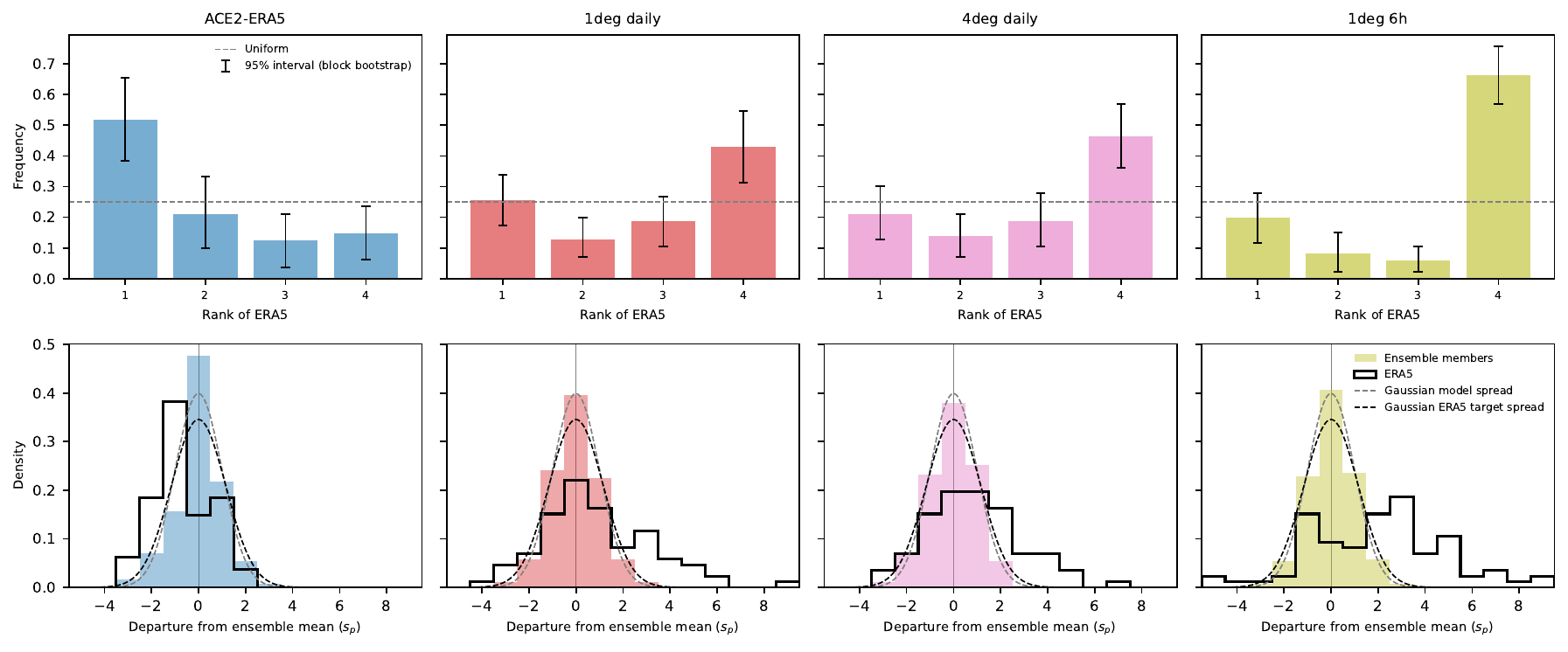}
\caption{Top: rank of ERA5's annual global-mean 2\,m temperature among the ensemble members for each year, over the AMIP runs of Figure~\ref{fig:climate_evolution}; a well-calibrated ensemble would produce a uniform distribution (dashed line). Comparison to the deterministic ACE2-ERA5 is possible because 3-member initial conditon ensembles are used. Error bars are the 95\% interval of each rank's frequency from a bootstrap over 5-year blocks of years with wraparound. ERA5 falls at the highest rank disproportionately often for the \ace models and at the lowest rank for ACE2-ERA5. Bottom: the same annual means expressed as departures from each year's ensemble mean, in units of the ensemble standard deviation pooled over all years ($s_p$), for the ensemble members (filled) and ERA5 (black outline). The member departures are scaled by $\sqrt{3/2}$ because each member contributes to its own year's mean, which makes the unit-variance Gaussian (grey dashed) their reference. ERA5 does not take part in the ensemble mean, so its departures are not scaled and are expected to be wider even for a calibrated ensemble: their reference is a Gaussian of variance $4/3$ (black dashed).}
\label{fig:ensemble_zscore}
\end{figure}

The ERA5 target values for 2-meter temperature and total water path are frequently well outside the envelope of the \ace ensemble variability (Figure~\ref{fig:ensemble_zscore}). The \ace models underestimate interannual variability, and in particular fail to represent positive extremes in the annual means. ACE2-ERA5 also underestimates interannual variability, but mostly misses negative extremes. The underestimation is robust to architecture changes. The calibration of \ace's ensemble at weather lead times (Section~\ref{sec:calibration}) therefore does not carry over to interannual timescales, and more work is needed before the ensemble spread can be interpreted as climate uncertainty.

\subsection{ENSO response}

\begin{figure}
\centering
\includegraphics[width=0.9\textwidth]{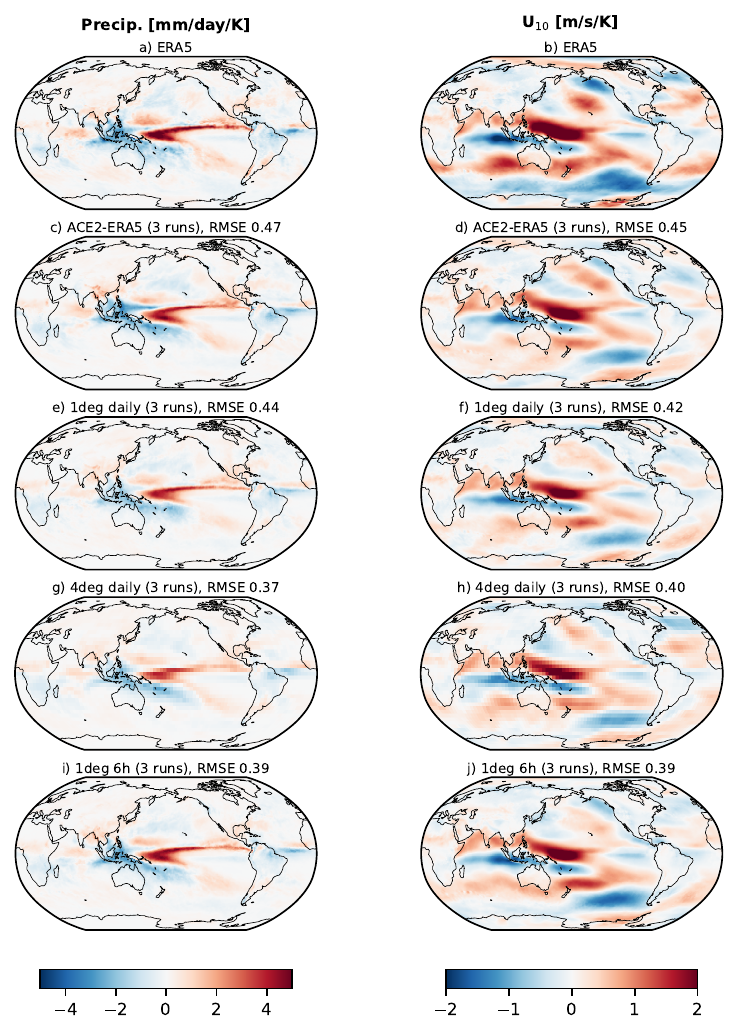}
\caption{Regression of monthly precipitation (left) and 10\,m zonal wind (right) on the Ni\~no-3.4 index over the 10-year runs, for ERA5 and the published models (averaged over initial conditions). Titles give the area-weighted pattern RMSE against ERA5, averaged over the per-run values.}
\label{fig:enso_maps}
\end{figure}

\begin{figure}
\centering
\includegraphics[width=0.8\textwidth]{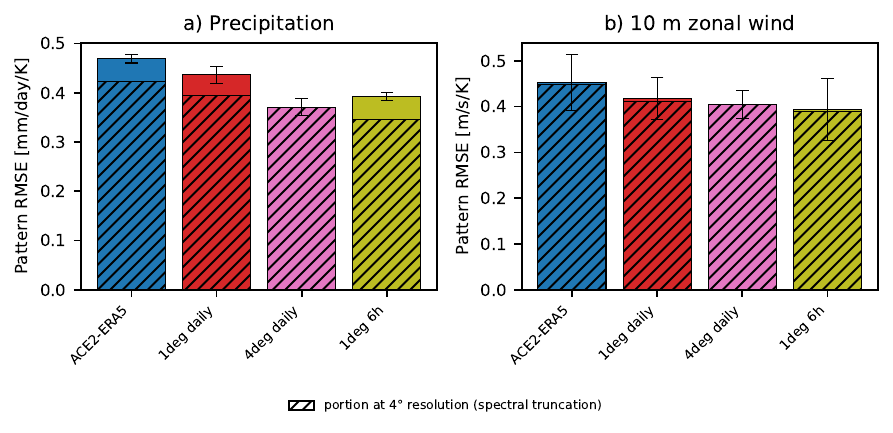}
\caption{Pattern RMSE of the Ni\~no-3.4 regression against ERA5 for precipitation (a) and 10\,m zonal wind (b), per model, each run scored separately then averaged. As in Figure~\ref{fig:time_mean_rmse}, the hatched lower segment is the RMSE after the error map is truncated to the 4$^\circ$ grid.}
\label{fig:enso_rmse}
\end{figure}

\ace and ACE2-ERA5 both place the central Pacific wet response and the Maritime Continent dry response of precipitation to the Ni\~no-3.4 index where ERA5 does (Figure~\ref{fig:enso_maps}). ACE2-ERA5's wet response is too broad and its Maritime Continent dry response too strong; both are improved in the \ace models, which on the other hand underestimate the strength of the tropical wet response. Scoring each run separately (Figure~\ref{fig:enso_rmse}a), \ace's pattern RMSE is 0.44\,mm/day/K against 0.47 for ACE2-ERA5, with 95\% intervals that nearly touch. \ace-4deg scores 0.37, below the 0.39 that \ace scores once truncated to the 4-degree grid. The \ace-6h model also slightly improves on the deterministic baseline, with a precipitation pattern RMSE of 0.39\,mm/day/K.

All models capture the equatorial westerly anomaly in 10\,m zonal wind (Figure~\ref{fig:enso_maps}, right column). \ace's pattern RMSE is lower than ACE2-ERA5's (0.42 against 0.45\,m/s/K), but the difference is smaller than the spread across initial conditions (Figure~\ref{fig:enso_rmse}b), so we do not claim an improvement in the wind response.

\subsection{Tropical cyclones}

\begin{figure}
\centering
\includegraphics[width=\textwidth]{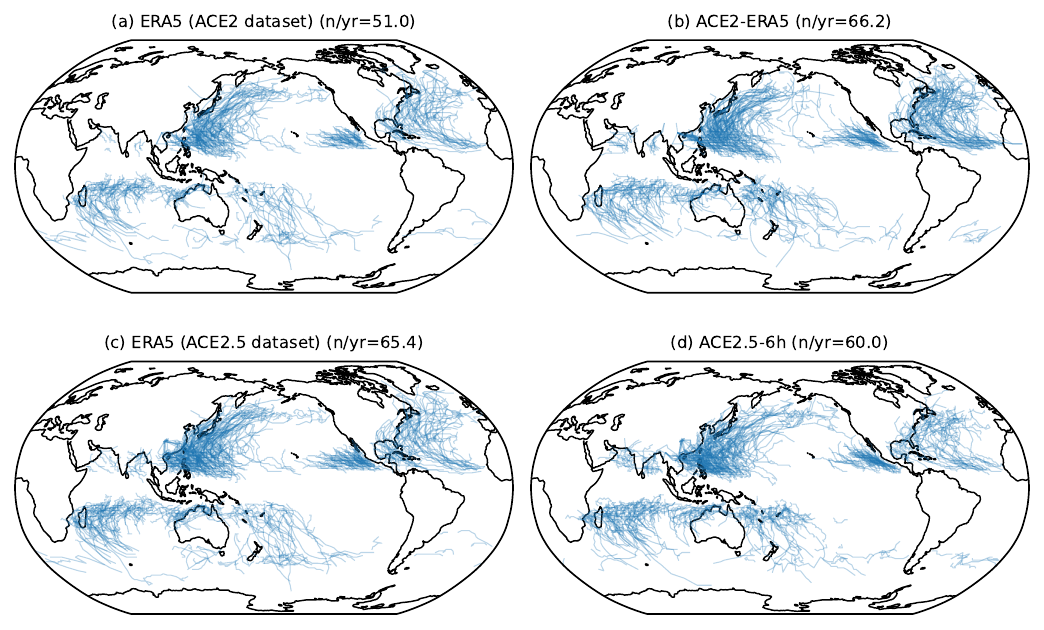}
\caption{Tropical cyclone tracks detected over the 10-year evaluation period (2001--2010). Each model is shown beside the ERA5 dataset it was trained on, because the two ERA5 datasets (differing in regridding method; Appendix~\ref{sec:si-data}) produce different track counts. Tracks are identified using TempestExtremes \citep{ullrich21} with the criteria of Appendix~\ref{sec:si-diagnostics}. Track counts per year are shown in each panel title.}
\label{fig:tc_tracks}
\end{figure}

Tropical cyclones need sub-daily output to track, so we evaluate them in \ace-6h, following \citet{wattmeyer25}: TempestExtremes \citep{ullrich21} detects and tracks storms in one 10-year run (2001--2010). Because the \ace and ACE2 ERA5 datasets use different regridding methods (xESMF conservative vs MetView interpolation; Appendix~\ref{sec:si-data}), TC detection on ERA5 itself gives different track counts: 65.4 storms per year on the \ace dataset versus 51.0 on the ACE2 dataset. This is due to tropical cyclone detection's reliance on identifying stable minima of sea-level pressure on a closed contour which persists for at least 54 hours. The bicubic interpolation used by the ACE2 dataset introduces noise in surface pressure and height which can disrupt this contour. We therefore compare each model against the ERA5 dataset it was trained on (Figure~\ref{fig:tc_tracks}). \ace-6h produces 60.0 storms per year, 8\% below its ERA5 target, while ACE2-ERA5 produces 66.2, 30\% above its own. Both models show realistic spatial activity in the western North Pacific, North Atlantic and South Indian Ocean basins. The intensity distributions are comparable between models (Figure~\ref{fig:tc_intensity}).

\subsection{Long-term stability}

\begin{figure}
\centering
\includegraphics[width=0.85\textwidth]{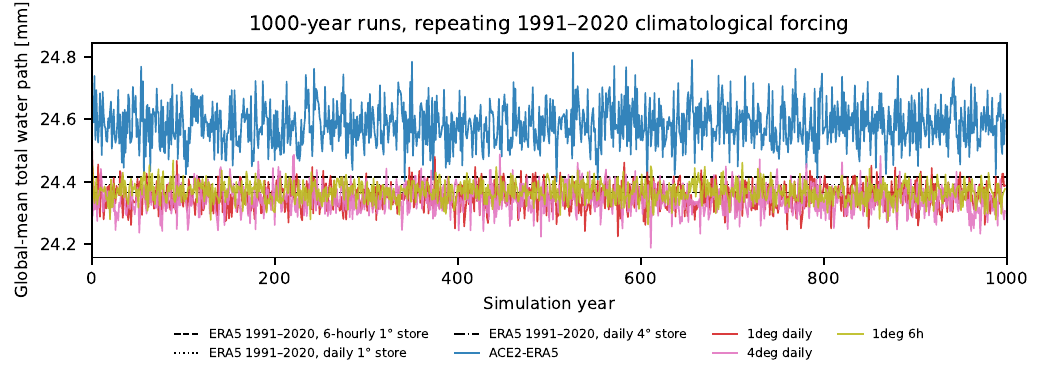}
\caption{Annual global-mean total water path over 1000-year runs forced by repeating 1991--2020 climatological sea surface temperature, sea ice, and CO$_2$.}
\label{fig:1000yr}
\end{figure}

\ace is stable over 1000 years under a repeating 1991--2020 climatological annual cycle of the prescribed forcings 
(Figure~\ref{fig:1000yr}). \ace settles near ERA5's 1991--2020 global-mean total water path of 24.39\,mm, ACE2-ERA5 is 0.17\,mm above the mean of its 6-hourly data, and \ace-6h is within 0.02\,mm of \ace. Runs are drift-free, with a linear trend in global-mean total water path below 0.05\,mm per 1000 years (0.2\% of the mean) in every run. The four trained seeds of \ace and \ace-4deg all produce stable 1000-year runs (Appendix~\ref{sec:seed-variability}).

\subsection{CO$_2$ sensitivity}

\begin{figure}
\centering
\includegraphics[width=\textwidth]{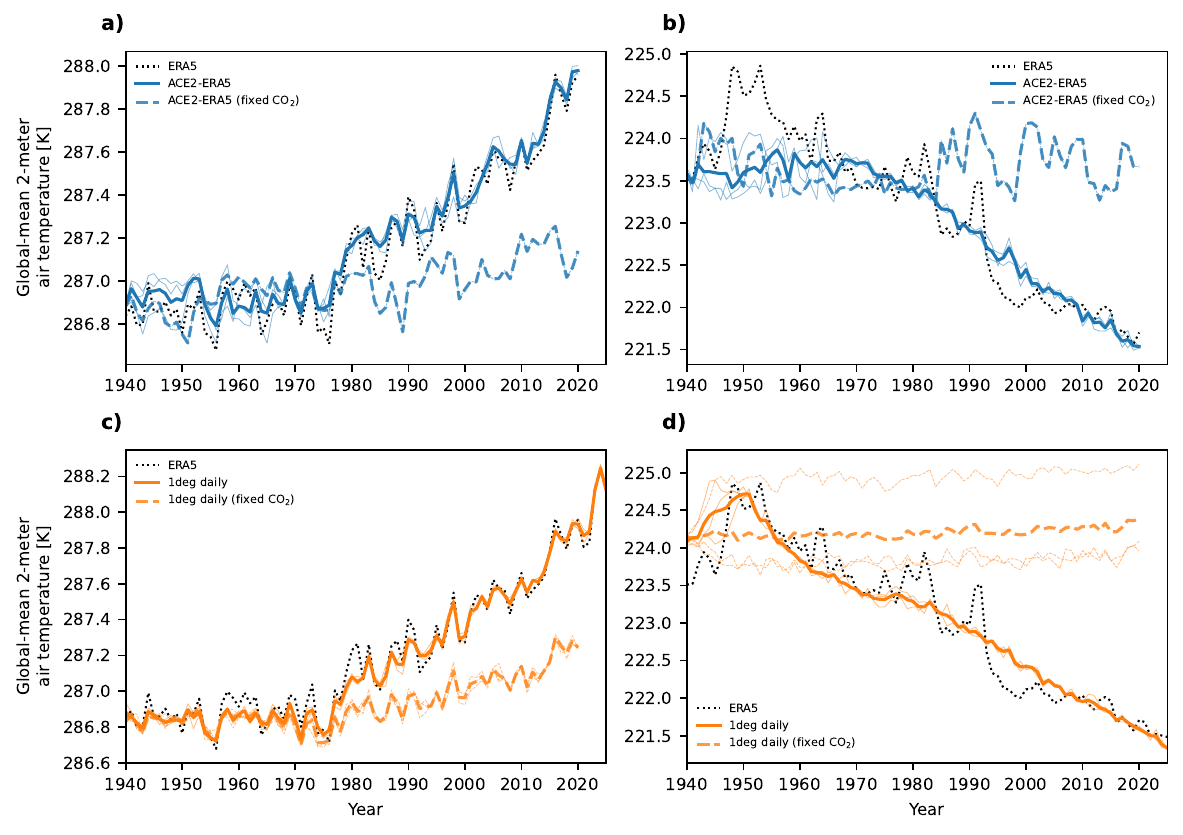}
\caption{Annual global-mean 2\,m temperature (a,~c) and stratospheric layer air temperature (b,~d) over the AMIP runs with historical CO$_2$ (solid) and fixed CO$_2$ held at the 1940 initial condition value (dashed), for ACE2-ERA5 (top) and \ace (bottom). ERA5 is shown as a dotted line. Thin lines are individual runs.}
\label{fig:co2_sensitivity}
\end{figure}

\ace takes global-mean CO$_2$ concentration as an explicit forcing input, and its AMIP runs use the observed CO$_2$ trajectory. To test whether the model has learned a physically plausible response to CO$_2$, we repeat the AMIP runs with CO$_2$ held constant at its 1940 value while all other forcings (SST, sea ice, solar radiation) evolve as observed.

Figure~\ref{fig:co2_sensitivity}a,c shows the global-mean 2\,m temperature. With fixed CO$_2$, the warming trend over 1980--2019 drops to roughly half of the ERA5 trend, even though the prescribed ocean SST is unchanged. This reduction is too large to be physical: land surface temperature should warm in response to the warming ocean even without CO$_2$ changes, yet the model predicts substantially cooler land surfaces under fixed CO$_2$, implying that it uses the CO$_2$ input as a shortcut for the forced warming signal rather than deriving the land response from the atmospheric circulation driven by SST. This is a known failure mode of ERA5-only training, present in both ACE2-ERA5 and \ace.

In the stratosphere (Figure~\ref{fig:co2_sensitivity}b,d), the response is more physically grounded: the observed cooling trend---driven primarily by CO$_2$---is absent in the fixed-CO$_2$ runs, confirming that the model has learned a qualitatively correct stratospheric CO$_2$ response.

\subsection{Ensemble calibration}
\label{sec:calibration}

\begin{figure}
\centering
\includegraphics[width=\textwidth]{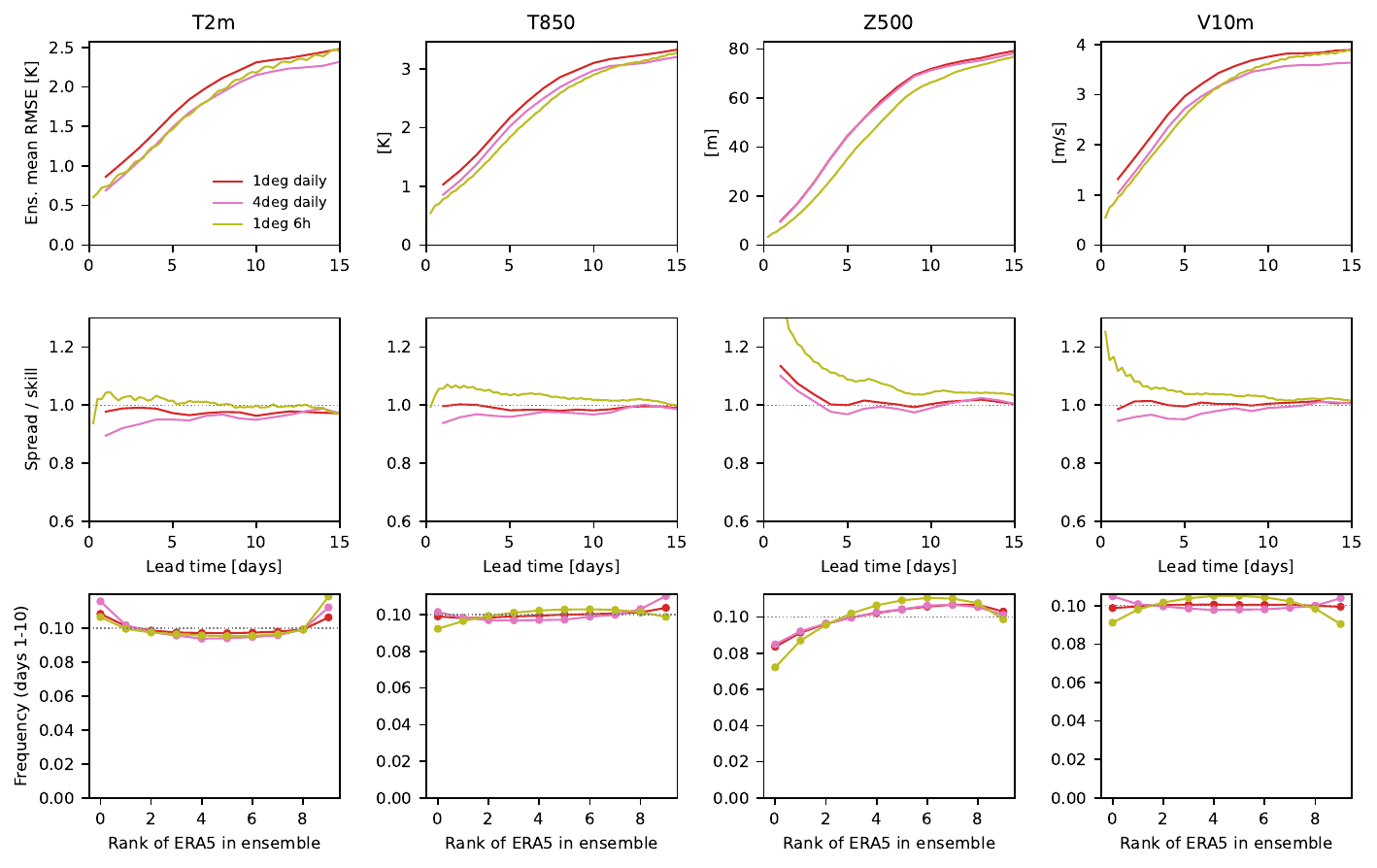}
\caption{Ensemble calibration of the 2020 weather forecasts for \ace, \ace-4deg and \ace-6h: ensemble spread and skill (top), their ratio (middle; 1 is perfectly calibrated), and the rank of the ERA5 value within the sorted 9-member ensemble, pooled over grid points, initial conditions and leads of 1--10 days (bottom). Skill is the RMSE of the ensemble mean corrected for finite ensemble size (Appendix~\ref{sec:si-ensemble-stats}).}
\label{fig:calibration}
\end{figure}

An ensemble is calibrated when its members are statistically indistinguishable from the truth. Two things then hold: the ensemble spread equals the error of the ensemble mean (a spread-to-skill ratio of 1), and the truth is equally likely to fall at any rank among the sorted members (a flat rank histogram). We test both on the 2020 weather forecasts (Figure~\ref{fig:calibration}).

\ace is well calibrated. Its spread-to-skill ratio is within 4\% of 1 from day 3 in every field, and between 0.97 and 1.00 at day 5. Its rank histograms are close to flat, with a small excess at the extreme ranks in 2\,m temperature. \ace-4deg is slightly under-dispersed (ratios of 0.94--1.02 from day 3). \ace-6h is slightly over-dispersed, most in 500\,hPa height, where the ratio is 1.15 at day 3 and 1.09 at day 5; in its other fields it is between 1.02 and 1.06 at those leads.

Calibration is not automatic for a model trained on an ensemble score. Both the spectral term of the loss and the spectral bottleneck contribute to it: without either one the ensemble is under-dispersed (Sections~\ref{sec:loss} and \ref{sec:bottlenecking}).

\subsection{Weather forecast skill}
\label{sec:weather}

\begin{figure}
\centering
\includegraphics[width=\textwidth]{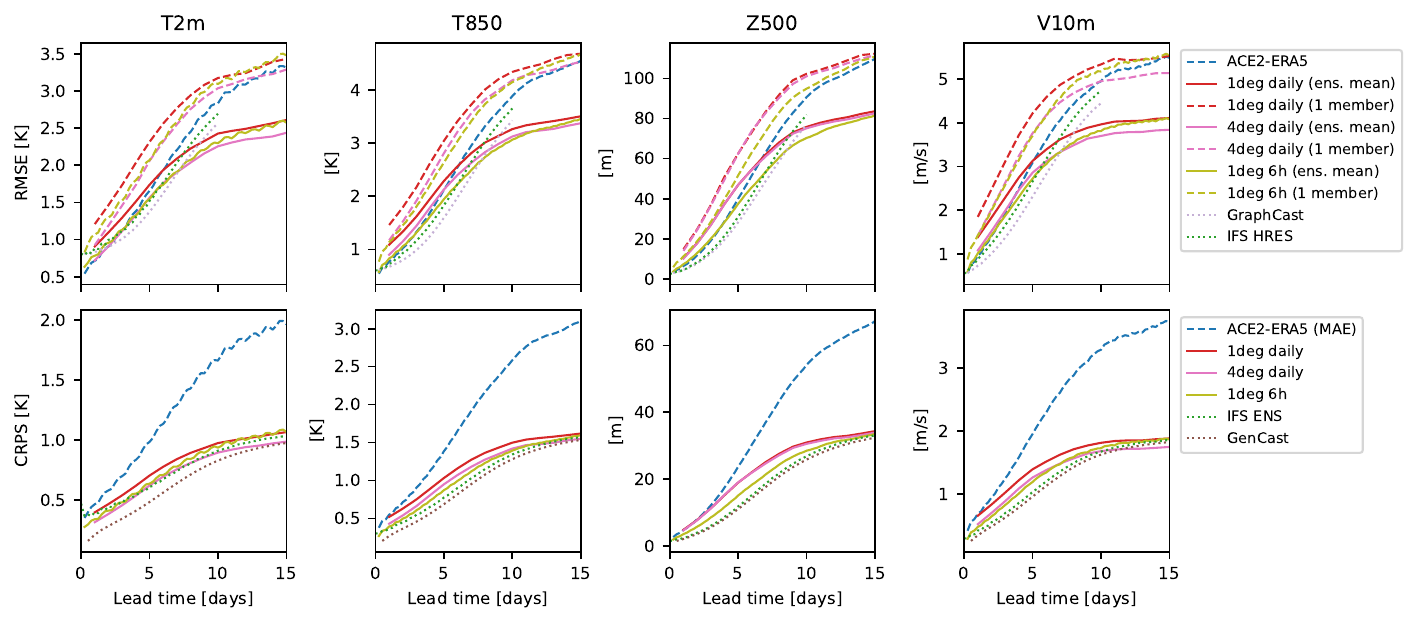}
\caption{Global RMSE (top) and CRPS (bottom) against ERA5 as a function of lead time, over 15-day forecasts from 52 weekly 00Z initial conditions in 2020. Solid lines are the 9-member ensemble mean (RMSE) and fair CRPS; dashed lines are a single member. For the deterministic ACE2-ERA5, CRPS reduces to mean absolute error.
Dotted lines are the WeatherBench2 2020 scorecards \citep{rasp24} for GraphCast and IFS HRES (RMSE) and IFS ENS and GenCast (CRPS), which average all 00Z and 12Z initial conditions of 2020 on a 1.5$^\circ$ grid; scoring those models on our weekly initial conditions instead moves their day-5 RMSE by at most 3\%.}
\label{fig:weather_skill}
\end{figure}

\ace is built for climate, and we have not tuned any configuration for weather skill. This section establishes how much skill the daily step gives up, and where the 6-hourly variant stands against weather models.

\paragraph{Reading a stochastic model's RMSE.} A single \ace member is less accurate than ACE2-ERA5 at every lead, with a day-5 500\,hPa height RMSE of 63\,m against 40\,m (Figure~\ref{fig:weather_skill}). This is expected of a calibrated ensemble, in which a member and the truth are independent draws from the same distribution, so a member's mean squared error is twice that of the mean of a large ensemble. The measured ratio of single-member to ensemble-mean RMSE at day 5 matches the theoretical value of 1.34 for 9 members to within 1\% (Appendix~\ref{sec:si-ensemble-stats}). In what follows we compare the 9-member ensemble mean with the deterministic ACE2-ERA5.

\paragraph{Daily against 6-hourly.} The daily step costs weather skill. At day 5, \ace's ensemble-mean RMSE is 5\% above ACE2-ERA5 in 2\,m temperature and 17\% above in 500\,hPa height, and it overtakes ACE2-ERA5 only after about a week. \ace-6h is 7\% and 6\% below ACE2-ERA5 in the same fields (1.55\,K and 37.5\,m). Most of that gap reflects training length rather than the time step itself. \ace-6h takes four times as many optimizer steps for the same number of epochs (511k vs 128k), and at a matched 128k steps the daily model's inline 500\,hPa height RMSE is 25\% lower than the 6-hourly model's (32 vs 42\,m). Appendix~\ref{sec:si-weather-gap} tests this and two other explanations.

\paragraph{Against weather models.} \ace-6h is the configuration intended for weather, so it is the one we compare (in Figure~\ref{fig:weather_skill}) with the WeatherBench2 baselines \citep{rasp24}. Its ensemble CRPS trails the IFS ensemble by 4--27\% at day 5 (least in 2\,m temperature, most in 500\,hPa height), by 3--6\% at day 10 and by 1--3\% at day 15. GenCast \citep{price25} is ahead of both throughout. The scorecards average all 732 initial conditions of 2020 on a 1.5-degree grid where we use 52, which changes the comparison little: rescoring GraphCast and IFS HRES on our initial conditions moves their day-5 RMSE by at most 3\%.

The gap at short leads should be read against training cost. GraphCast \citep{lam23} was trained for about 180 passes over 39 years of 6-hourly ERA5 (311,000 gradient updates at a batch size of 32), and GenCast for about 2,200 passes over 40 years of 12-hourly data at 1 degree followed by 70 more at 0.25 degrees (2.06 million updates at a batch size of 32). \ace sees each of its 70 years of daily data 50 times (160,000 updates at a batch size of 8). Training longer would improve weather skill, but we stop here because extended pre-training degraded the multi-decadal climate of some seeds (Appendix~\ref{sec:si-weather-gap}). We also avoid residual prediction, which improves the weather skill of ACE but often leads to large biases in long rollouts (result not shown).

\subsection{Training reduced resolution models for climate}
\ace-4deg requires an order of magnitude fewer GPU hours to train than \ace (Table~\ref{tab:throughput}), yet reaches comparable climate skill at its target resolution. The evidence is the hatched bars of Figures~\ref{fig:time_mean_rmse} and \ref{fig:enso_rmse}, which truncate \ace's errors to the 4-degree grid: \ace-4deg's RMSE is close to them in 2\,m temperature, total water path and 500\,hPa height, and lower in precipitation and in the precipitation response to ENSO. In the AMIP runs it tracks the 2\,m temperature record at least as closely as \ace and repeats \ace's total water path offsets (Figure~\ref{fig:climate_evolution}). Training at reduced resolution therefore lets us iterate quickly on new ideas and run multi-decadal evaluations frequently during development, though verification at 1-degree resolution is still necessary and result transfer is not guaranteed. One still unexplored idea is to first train a coarse resolution climate emulator, and then train successively higher resolution emulators whose output is conditioned on the output of the frozen coarse emulator, as also proposed by \citet{witte25}. This one-way dependence should enable high-resolution simulations whose climate skill (at coarse scales) is determined by the coarse-resolution emulator. Additionally, the applicability of reduced resolution training is not yet determined for coupled emulator models.

\section{Model architecture and design}
\label{sec:architecture}

\begin{figure}
\centering
\includegraphics[width=\textwidth]{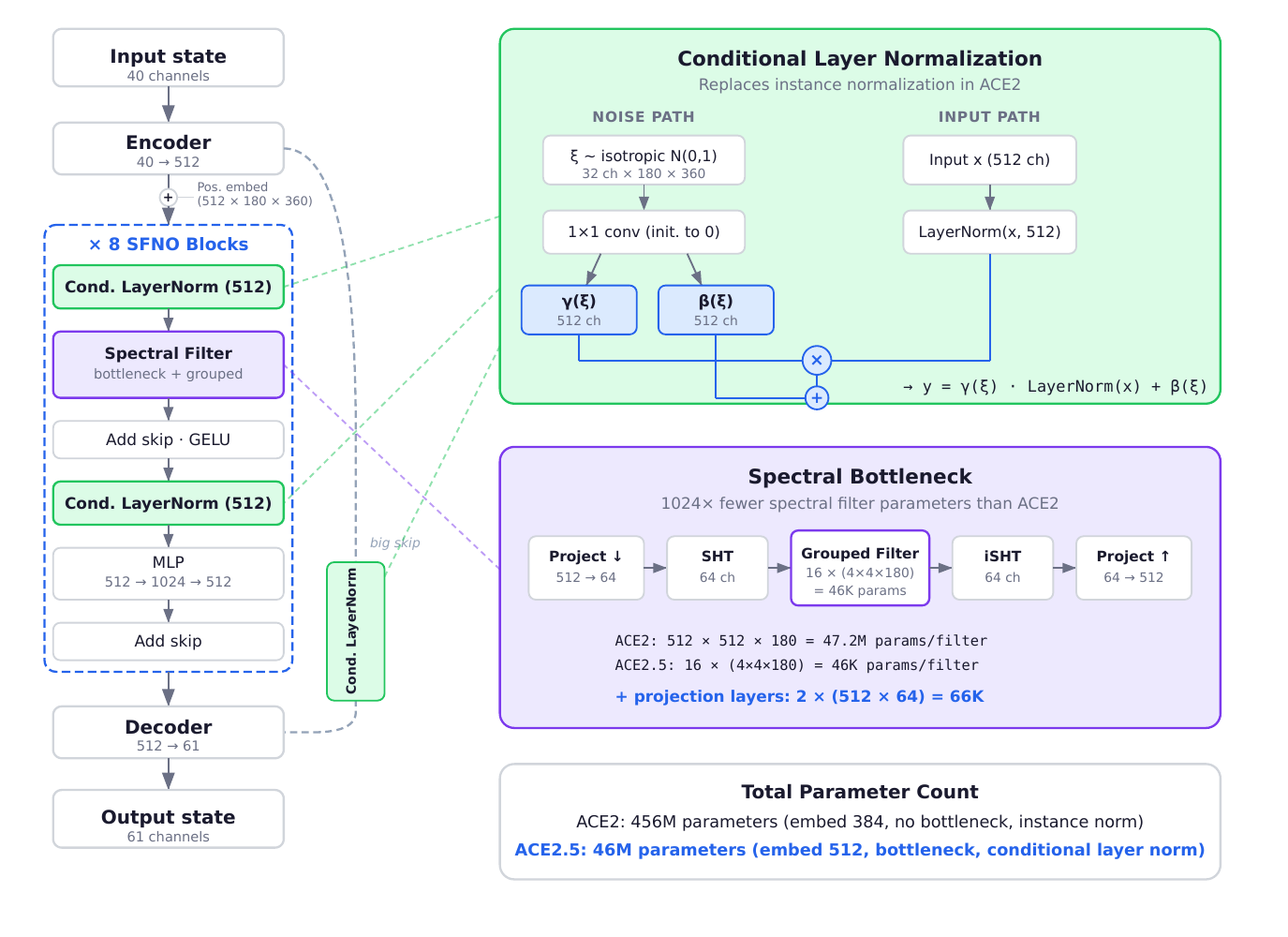}
\caption{Schematic of the \ace architecture. Left: the encoder--decoder backbone with 8 repeated SFNO blocks and a skip connection from the encoder to the decoder. Green boxes are conditional layer normalization layers (detailed top right), which replace ACE2's instance normalization; the purple spectral filter box uses the spectral bottleneck (detailed bottom right). Grey components are shared with ACE2.}
\label{fig:architecture}
\end{figure}

\ace is a modified version of ACE2 \citep{wattmeyer25}, a deterministic Spherical Fourier Neural Operator (SFNO; \citealp{bonev23}) with constraints that conserve dry air mass and moisture. It differs from ACE2 in that it is stochastic, in the size of its spectral filters, and in its inputs and outputs. The architecture is shown schematically in Figure~\ref{fig:architecture}.

\paragraph{Noise conditioning.} Following AIFS-CRPS \citep{lang24}, GenCast \citep{price25} and FGN \citep{alet25}, we replace ACE2's instance normalization with conditional layer normalization. Layer normalization normalizes features across channels independently for each column, instead of independently across the horizontal domain as in \citet{wattmeyer23} and \citet{wattmeyer25}. The scale and bias of each normalization layer are affine functions of 32 noise channels, drawn independently for every forward step as isotropic Gaussian white noise on the sphere (unit variance, flat spherical harmonic spectrum), as in \citet{price25}. Each noise channel maps to a per-column scale and bias through a learned $1 \times 1$ convolution, initialized to zero so that the model starts training as a deterministic network. The conditioning is applied at both normalization layers of each of the 8 SFNO blocks and at the normalization of the skip connection from the encoder to the decoder. \ace shares roots with FourCastNet, especially the use of SFNO blocks \citep{bonev23}, but its conditional normalization and training methodology are distinct from those of FourCastNet 3 \citep{bonev25}.

\paragraph{Width and size.} The embedding dimension is 512 features, up from 384 in ACE2. Changes to the spectral filters (Section~\ref{sec:bottlenecking}) nonetheless leave a model of 46 million parameters, compared with 456 million for ACE2-ERA5.

\ace takes 40 inputs: the prognostic state (surface pressure, surface temperature, and air temperature, specific total water and horizontal winds on 8 terrain-following layers) and the forcings (land, ocean and sea ice fractions, incoming solar radiation at the top of the atmosphere, surface topography and global-mean CO$_2$ concentration), with sea surface temperature prescribed over ocean. Unlike ACE2, the 2\,m temperature and specific humidity and the 10\,m winds are diagnostic outputs only rather than prognostic inputs, reducing the input count by four. To the outputs of ACE2 we add mean sea level pressure, the surface wind stresses and the frozen precipitation rate, and 7 diagnostic outputs for daily mean near-surface fields (skin temperature, surface pressure, 2\,m temperature and specific humidity, the 10\,m wind components and 10\,m wind speed), each computed as the average of the 24 hourly values ending at and including the output snapshot. The dry air mass and moisture budget corrections, and the clipping of non-negative outputs, follow ACE2; the frozen precipitation rate is additionally clipped to at most the total precipitation rate. Table~\ref{tab:fields} lists all input and output fields.

\begin{table}
\centering
\caption{Input and output fields of \ace (daily timestep). Fields marked with $\dagger$ are new relative to ACE2-ERA5. Vertically resolved fields span 8 terrain-following layers. Diagnostic outputs are predicted but not fed back as inputs. The 6-hourly model (\ace-6h) omits the daily-mean diagnostics.}
\label{tab:fields}
\small
\begin{tabular}{llc}
\toprule
Field & Role & Levels \\
\midrule
\multicolumn{3}{l}{\emph{Inputs only (prescribed forcing)}} \\
Land, ocean, sea ice fractions & Static/prescribed & 1 each \\
Incoming solar radiation (TOA) & Prescribed & 1 \\
Surface topography & Static & 1 \\
Global-mean CO$_2$ & Prescribed & 1 \\
\midrule
\multicolumn{3}{l}{\emph{Prognostic (input and output)}} \\
Surface pressure & Prognostic & 1 \\
Surface temperature & Prognostic & 1 \\
Air temperature & Prognostic & 8 \\
Specific total water & Prognostic & 8 \\
Eastward wind & Prognostic & 8 \\
Northward wind & Prognostic & 8 \\
\midrule
\multicolumn{3}{l}{\emph{Diagnostic outputs}} \\
2\,m temperature, specific humidity & Diagnostic & 1 each \\
10\,m zonal and meridional wind & Diagnostic & 1 each \\
850\,hPa temperature & Diagnostic & 1 \\
500\,hPa geopotential height & Diagnostic & 1 \\
Surface precipitation rate & Diagnostic & 1 \\
$\dagger$ Frozen precipitation rate & Diagnostic & 1 \\
$\dagger$ Mean sea level pressure & Diagnostic & 1 \\
$\dagger$ Eastward, northward surface stress & Diagnostic & 1 each \\
Radiative fluxes (LW/SW, up/down, sfc/TOA) & Diagnostic & 6 \\
Latent and sensible heat flux & Diagnostic & 1 each \\
Water path advection tendency & Diagnostic & 1 \\
\midrule
\multicolumn{3}{l}{\emph{$\dagger$ Daily-mean diagnostic outputs}} \\
Skin temperature, surface pressure & Diagnostic & 1 each \\
2\,m temperature, specific humidity & Diagnostic & 1 each \\
10\,m zonal wind, meridional wind, wind speed & Diagnostic & 1 each \\
\bottomrule
\end{tabular}
\end{table}

\subsection{Training}
\label{sec:training}

All \ace models except \ace-6h are trained on ERA5 at a 24-hour time step, on the years 1940--1995, 2011--2019 and 2021--2025, withholding 1996--2010 and 2020 for evaluation; 1996--1997 serve as the validation set used to select checkpoints. This matches the ACE2-ERA5 training years (which included 2021--2022) except for the addition of 2023--2025. The 1-degree and 4-degree datasets are new regriddings of ERA5 from the native 0.25$^\circ$ resolution to Gaussian grids using area-conservative regridding (xESMF; Appendix~\ref{sec:si-data}), replacing the MetView-based regridding of the ACE2 dataset \citep{wattmeyer25}. The daily datasets are produced by time-coarsening the 6-hourly data: snapshot fields retain the 00Z value, and window fields (fluxes, precipitation) are averaged over the four 6-hourly values. Inputs and targets are normalized as in \citet{wattmeyer25}: prognostic outputs by residual scaling, diagnostic outputs by full-field scaling, with statistics computed from the daily (or 6-hourly, for \ace-6h) ERA5 data over 1990--2019.

The model is pre-trained for 40 epochs on 1 forward step with 2 ensemble members per sample. Fine-tuning starts from the exponential-moving-average (EMA) weights at the final pre-training epoch; we did not find a meaningful difference when fine-tuning from the non-EMA weights instead.

The weights for \ace and \ace-4deg are then fine-tuned for 10 epochs on 3 forward steps with gradients detached between forward steps. \ace-6h is fine-tuned for 2 epochs on 12 forward steps, so that it also corresponds to 3 days. Note this differs from \citet{perkins26, wu26, clark2026}, which optimize on the last step of a stochastically varying number of forward steps. Our results are not sensitive to this choice: a variant of \ace-6h trained with the stochastically varying schedule of \citet{perkins26} (\ace-6h-svfs) achieves similar weather skill and comparable climate fidelity (Appendix~\ref{sec:si-loss-ablation}), and the fixed schedule is simpler.

Detaching gradients has the benefit of keeping the fine-tuning gradients closer to those seen for 1-step pretraining, as the model is effectively being optimized over three separate single forward steps. It also avoids the possibility that the network will intentionally produce an incorrect prediction on an earlier step in order to improve skill on later steps. We found when training deterministic ACE2-ERA5 over two forward steps, small but robust banding artifacts would appear in the 1-step evolution of the humidity layer second from the top of atmosphere, which went away when detaching gradients between steps (result not shown). Detaching gradients between forward steps also allows us to apply gradient accumulation, requiring only a single step of gradients be held in memory at a time.

Both stages of training use the AdamW optimizer with a constant learning rate of $10^{-4}$ and weight decay of 0.01, a batch size of 8 (one sample per GPU on 8 H100 GPUs at 1 degree; a single GPU at 4 degrees), and an EMA of the weights with decay 0.999. The EMA weights are used for validation and for every evaluation in this paper: the evaluated checkpoint is the EMA weights at fine-tuning epoch 10 (epoch 2 for \ace-6h). Over the 25,562 training samples, the two stages amount to about 160,000 gradient updates; Section~\ref{sec:weather} compares this with the training of GraphCast and GenCast.

\ace-4deg is trained at 4-degree resolution on the same years of ERA5, coarsened to 4 degrees, with the same architecture and loss. \ace-6h uses a 6-hour forward timestep on the 6-hourly ERA5 dataset; its outputs are those of \ace without the seven daily-mean diagnostics. \ace-6h sees four times as many gradient updates per epoch during pre-training as the daily variants, and takes four times as long to pre-train as \ace.

\subsection{Loss function}
\label{sec:loss}

Models trained with CRPS alone produce excess spectral power at small scales \citep{alet25, swift}. Three remedies have been used: AIFS-CRPS \citep{lang24} filters small scales in the residual architecture, FourCastNet 3 \citep{bonev25} adds the CRPS of the spectral power as a loss term, and NeuralGCM \citep{kochkov24} adds the energy score of the complex spherical harmonic coefficients (a spectral CRPS) at wavenumbers below 80.

We take the third approach. Our loss function (Appendix~\ref{eq:crps-definition}) combines a nodal CRPS with a spectral energy score, and differs from NeuralGCM's in three ways: no wavenumber cutoff, no area weighting on the nodal CRPS, and fair (rather than almost-fair) CRPS, evaluated with 2 ensemble members. We scale the energy score by an empirical factor of $1/(2\sqrt{N_\ell N_m})$ (Equation~\ref{eq:energy-score}), where $N_\ell = N_m = 180$ are the numbers of retained degrees and orders in the spectral space. This gives it a magnitude similar to the CRPS when both are evaluated on Gaussian-distributed random variables, regardless of domain size; modes with $m > 0$ are counted twice, for their conjugate pair. We then empirically choose weights of $0.9$ on the nodal CRPS and $0.1$ on the spectral energy score, as in \citet{perkins26}. Losses are computed on variables normalized as in \citet{wattmeyer25}, but without variable-specific weights, as these were chosen in ACE2-ERA5 specifically to deal with high loss values in stochastic fields when using a deterministic loss function. We did retain a 5$\times$ weight for 500\,hPa height, which was necessary to avoid underfitting this field. This loss function was also used in the stochastic ACE models of \citet{perkins26}, \citet{wu26} and \citet{clark2026}, except that their models inherited the variable-specific weights from ACE2-ERA5.

\begin{figure}
\centering
\includegraphics[width=\textwidth]{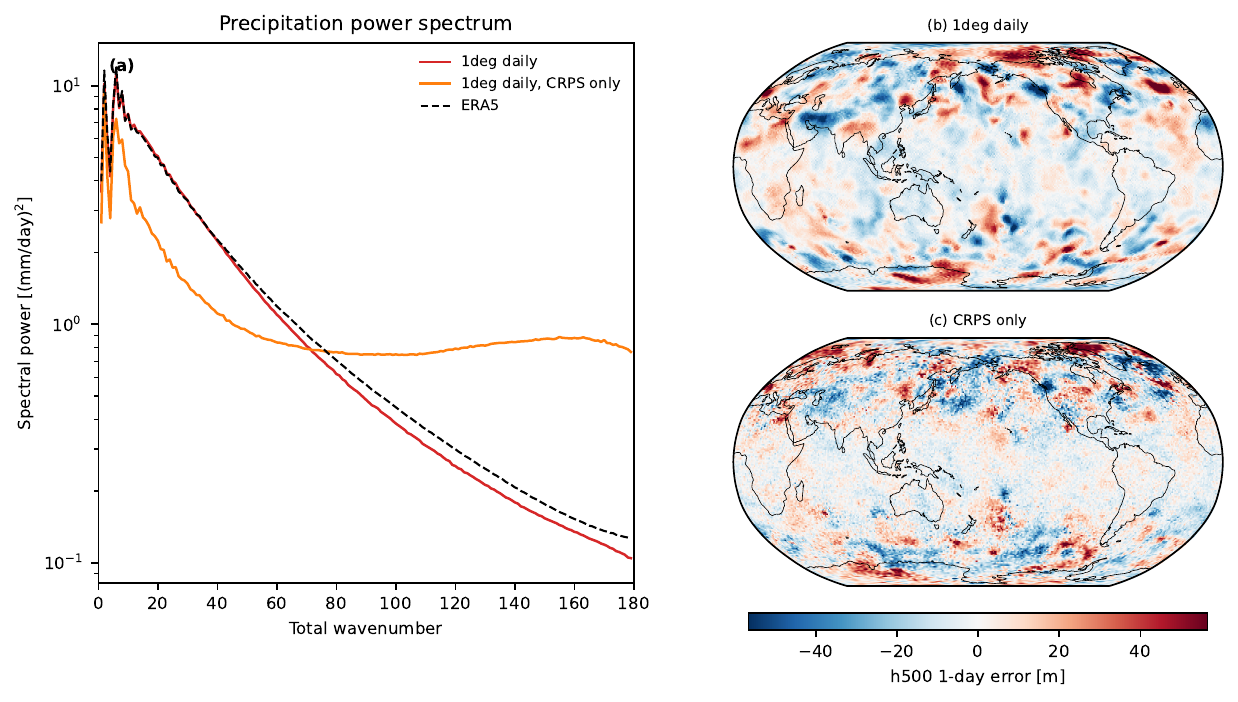}
\caption{Effect of the spectral energy score loss term. (a) Mean spherical power spectrum of daily-mean surface precipitation rate for \ace, \ace-crps-only (trained with CRPS loss alone, no spectral energy score) and ERA5. (b, c) 500\,hPa geopotential height 1-day prediction error (prediction minus ERA5) from a single forecast for \ace (b) and \ace-crps-only (c). Training without the spectral term produces excess small-scale noise in the predictions.}
\label{fig:loss_motivation}
\end{figure}

To demonstrate the importance of the spectral term, we train an ablation model (\ace-crps-only) with pointwise CRPS alone, retaining the 5$\times$ weight for 500\,hPa height. Figure~\ref{fig:loss_motivation} shows the result: \ace-crps-only produces excess spectral power at small scales in precipitation and visibly noisier 500\,hPa height predictions. This noise occurs in all fields with significant stochastic variability, such as stratospheric temperature, and pollutes all fields during autoregressive rollout. The spectral energy score suppresses this noise without filtering. The effect extends to the climate: \ace-crps-only's time-mean (daily-mean) 2\,m temperature RMSE is 0.94\,K against 0.20\,K for \ace, and its precipitation RMSE is 0.65\,mm/day against 0.34 (Figure~\ref{fig:si_loss_rmse}). CRPS-only training also produces an under-dispersed ensemble, with the spread-to-skill ratio falling to 0.89--0.93 by day 5 and to 0.75--0.89 by day 15 (Figure~\ref{fig:si_loss_calibration}).

\subsection{Spectral bottlenecking}
\label{sec:bottlenecking}

Most of the parameters of an SFNO are in its spectral filters (Figure~\ref{fig:architecture}, purple block). A filter has shape \texttt{[embed\_dim, embed\_dim, l\_values]}: it defines a separate linear map between features for each total wavenumber $l$, of which there are 180 at 1-degree resolution. By contrast, the MLP blocks have filters of shape \texttt{[embed\_dim, embed\_dim]}. We shrink the spectral filters in two ways.

\paragraph{Spectral bottleneck.} A parameter \texttt{spectral\_ratio} sets the width of the spectral filters relative to the rest of the network. We use 1/8, which reduces the spectral filter parameter count by a factor of 64 and brings it to the same order of magnitude as the local MLP filter. The bottleneck is a linear projection from the full 512-feature latent space to a 64-feature space before the spherical harmonic transform (SHT), and a projection back after the filter and the inverse SHT. Because each operation is linear, this is identical to applying the bottleneck in spectral space, with the added benefit of reducing SHT workload. The ratio was tuned to minimize validation loss on an earlier version of the model at 4-degree resolution, and was not re-tuned at 1 degree.
Ablations at 1-degree resolution confirm this choice transfers: ratios of 1/2, 1/4, 1/8 and 1/16 all produce 10-year time-mean 2\,m temperature RMSEs within 0.02\,K of each other and weather forecast RMSEs within 3\%, so the 1/8 ratio sits at the knee of the parameter-count curve with no skill penalty (Appendix~\ref{sec:si-loss-ablation}).

\paragraph{Grouped convolutions.} Following the Makani implementation of FourCastNet 3 \citep{bonev25}, we partition the spectral filter into a block-diagonal layout of $N$ filters of shape \texttt{[in\_channels / N, out\_channels / N, l\_values]}, reducing the parameter count by a further factor of $N$. FourCastNet 3 uses groups to isolate input and output fields; we use them only to reduce parameter count, and let the network partition information between groups as it wishes. We use $N = 16$.

\begin{figure}
\centering
\includegraphics[width=\textwidth]{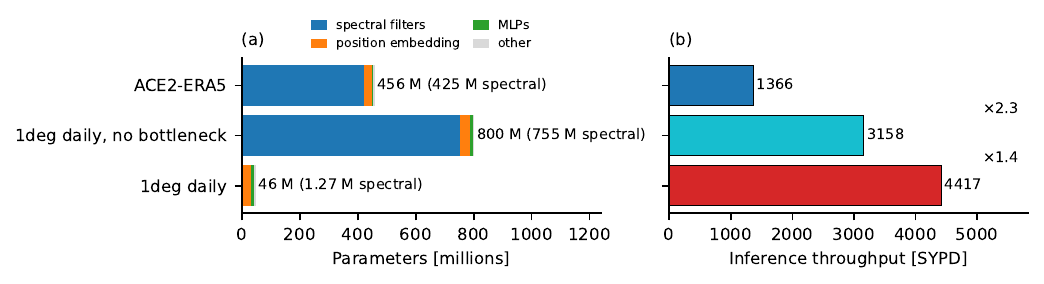}
\caption{The two steps from ACE2-ERA5 to \ace. (a) Trainable parameters by component; the spectral filters include the bottleneck's projections. (b) Inference throughput in simulated years per day on one H100 GPU (Table~\ref{tab:throughput}), with the ratio between successive models. ``1deg daily, no bottleneck'' is \ace-no-bottleneck, which differs from ACE2-ERA5 in its daily time step, wider latent dimension, noise conditioning, loss and variable set.}
\label{fig:cost}
\end{figure}

\paragraph{Resulting size.} At 1-degree resolution each spectral filter is 16 filters of shape \texttt{[4, 4, 180]}, or 46,080 complex parameters, a factor of 1024 fewer than the 47,185,920 a 512-wide filter would need without these two changes. Counting the bottleneck's two projection matrices as well (64$\times$512 in and 512$\times$64 out, 65,536 real weights against 92,160 for the 16 complex filters), the spectral convolution as a whole shrinks by a factor of about 600 relative to a 512-wide unbottlenecked filter, and about 340 relative to ACE2-ERA5's 384-wide filter. The model as a whole has 46 million parameters, against 800 million for the same architecture without the two changes (\ace-no-bottleneck) and 456 million for ACE2-ERA5 (Figure~\ref{fig:cost}a). The largest component that remains is the learned position embedding, at 33 million parameters.

\paragraph{Effect on cost.} Figure~\ref{fig:cost}b separates the two contributions to \ace's inference throughput, which is 3.2$\times$ that of ACE2-ERA5. Without the bottleneck, the daily model runs 2.3$\times$ faster than ACE2-ERA5: it takes a quarter as many steps, but each step costs more because of the wider latent dimension and the noise conditioning. The bottleneck yields a further 1.4$\times$ speedup (4417 vs 3158 SYPD). The bottleneck helps training more than inference: fine-tuning has a throughput of 1.43 samples per second per GPU (3 forward steps, batch size 8 on 8 GPUs), against 0.90 for the no-bottleneck model --- a 1.6$\times$ speedup. Training is also shorter than ACE2-ERA5's in absolute terms, at 193 GPU-hours against 914: \ace's 50 epochs of daily data are 4.6$\times$ fewer gradient updates than ACE2-ERA5's 120 epochs of 6-hourly data. Table~\ref{tab:throughput} gives the training cost and inference throughput of every configuration. \ace uses about 10\,GB of GPU memory during inference when buffering 40 forward steps for data loading and the output aggregators; this can be reduced at the cost of potentially slower inference.

\begin{table}
\centering
\caption{Model size, training cost and inference throughput. Parameters are trainable parameters in millions. \ace-deterministic and \ace-no-bottleneck are ablations (Table~\ref{tab:ablations}). Training GPU-hours are wall time at the checkpoint epoch (including validation and inline evaluation) times GPU count; the pretrain and fine-tune are on 8 H100 GPUs for 1-degree models and 1 GPU for 4-degree. Inference throughput is simulated years per wall-clock day (SYPD) on a single H100 GPU with 8 ensemble members and no I/O, counting all members (mean of 3 replicates; the scatter across replicates is a few percent).}
\label{tab:throughput}
\small
\setlength{\tabcolsep}{4pt}
\begin{tabular}{lrrrrrr}
\toprule
Model & Params & \multicolumn{2}{c}{Pre-train} & \multicolumn{2}{c}{Fine-tune} & Inference \\
\cmidrule(lr){3-4} \cmidrule(lr){5-6}
 & (M) & Epochs & GPU-h & Epochs & GPU-h & SYPD \\
\midrule
\ace & 46 & 40 & 129 & 10 & 64 & 4417 \\
\ace-6h & 46 & 40 & 480 & 2 & 189 & 1177 \\
\ace-4deg & 14 & 40 & 12 & 10 & 6 & 22897 \\
\ace-no-bottleneck & 800 & 40 & 213 & 10 & 106 & 3158 \\
\ace-deterministic & 46 & 40 & 123 & 10 & 39 & 4991 \\
ACE2-ERA5 \citep{wattmeyer25} & 456 & 120 & 914 & --- & --- & 1366 \\
\bottomrule
\end{tabular}
\end{table}

\begin{figure}
\centering
\includegraphics[width=\textwidth]{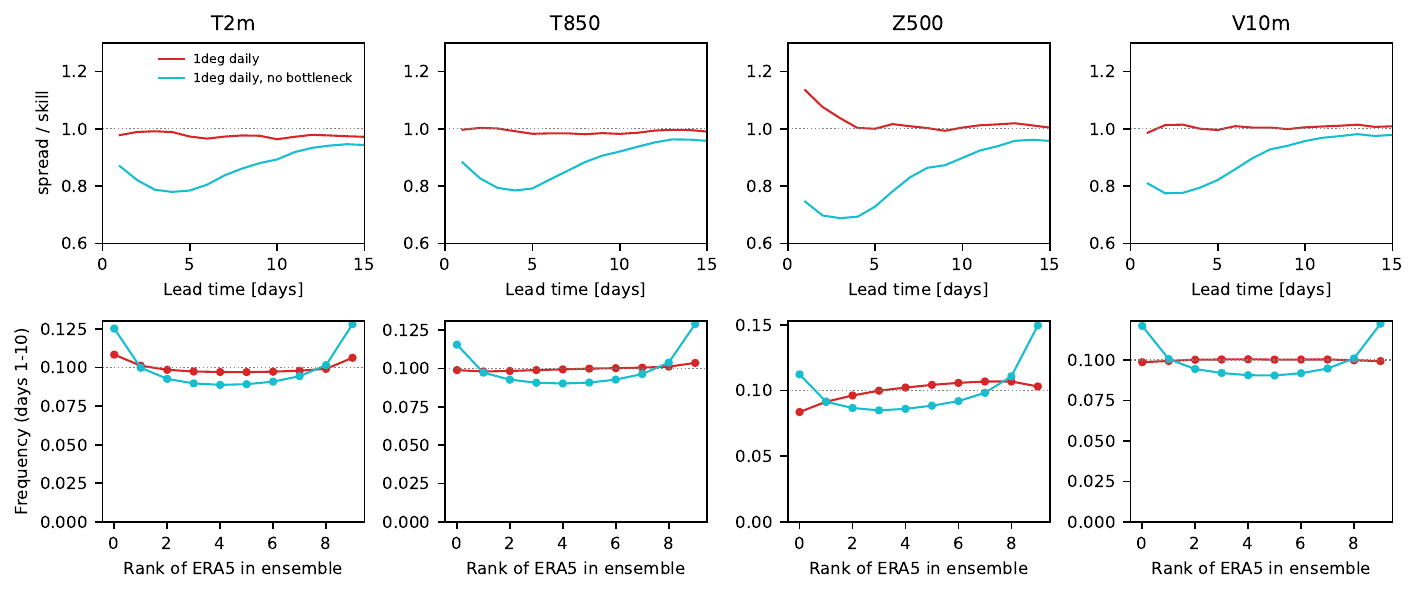}
\caption{Effect of the spectral bottleneck on ensemble calibration. Top: ratio of ensemble spread to skill (RMSE of the ensemble mean corrected for finite ensemble size) over the 2020 weather forecasts for \ace and \ace-no-bottleneck (the same model without the spectral ratio or grouped convolutions); a ratio of 1 is perfectly calibrated. Bottom: rank of the ERA5 value within the sorted 9-member ensemble, pooled over leads of 1--10 days.}
\label{fig:bottleneck_calibration}
\end{figure}

\paragraph{Effect on calibration.} Beyond reducing cost, the bottleneck improves ensemble calibration. \ace-no-bottleneck uses a \texttt{spectral\_ratio} of 1 and a single group, restoring the full spectral filter. Figure~\ref{fig:bottleneck_calibration} compares its calibration with that of \ace. \ace-no-bottleneck is under-dispersed by 12--31\% over forecast days 1--5 (spread-to-skill ratio 0.69--0.88) and its rank histograms are U-shaped, with ERA5 falling outside the ensemble 11--15\% of the time at each end against the 10\% of a uniform histogram. It recovers to within about 6\% of 1 by day 15. \ace, with the bottleneck, is within 4\% of 1 from day 3 in every field (Section~\ref{sec:calibration}). The bottleneck thus recovers short-lead ensemble spread. This is likely the result of the no-bottleneck model overfitting on the training dataset with more parameters, and it is possible predictions could be calibrated for a model stopped earlier in training (untested). However this would worsen weather prediction skill, and it is notable that the validation loss was decreasing when training stopped.

\section{Conclusions}

\ace replaces ACE2-ERA5's deterministic architecture with noise-conditioned layer normalization trained on a combined CRPS and spectral energy score loss. The result is a calibrated ensemble: in weather forecasts its spread matches its error to within 4\% from day~3 in every field, with day-5 spread-to-skill ratios of 0.97--1.03 across the four trained seeds. It also recovers the grid-scale precipitation structure that deterministic training suppresses, retaining 86\% of its ERA5 target's spectral power at wavenumbers 100--179 where ACE2-ERA5 retains 36\%, and reproducing the frequency of extreme precipitation events.

These improvements come at lower computational cost. The spectral bottleneck and grouped convolutions leave \ace a tenth the size of ACE2-ERA5, and together with the daily forward step give it 3.2$\times$ the inference throughput. Climate skill is consistent across seeds, so a single training run suffices.

On climate timescales, \ace tracks the observed 1940--2020 warming with a global-mean 2\,m temperature error of 0.05\,K, improves the precipitation response to ENSO relative to ACE2-ERA5, and is stable for 1000 years under repeating climatological forcing. The 4-degree variant (\ace-4deg) reaches comparable climate skill at its target resolution while training in an order of magnitude fewer GPU hours, making reduced-resolution training a practical tool for rapid development. The 6-hourly model (\ace-6h) recovers the ensemble-mean weather skill of its deterministic predecessor while adding calibrated spread.

Several limitations remain.
\begin{itemize}
\item \emph{Interannual variability is underestimated.} ERA5's annual global-mean 2\,m temperature falls outside the ensemble envelope significantly more often than expected under a well-calibrated climate ensemble, and the rank histograms confirm this is not a sampling artifact. More work is needed before the ensemble spread can be interpreted as climate uncertainty.
\item \emph{The surface response to CO$_2$ is not physical.} Like ACE2-ERA5, \ace uses its CO$_2$ input as a shortcut for the forced warming of the land surface: with CO$_2$ held fixed, the 1980--2019 trend in global-mean 2\,m temperature drops to roughly half of the ERA5 trend even though sea surface temperature is prescribed. The model should not be used to separate the effects of CO$_2$ and sea surface temperature, or with combinations of the two that are far from the historical record.
\item \emph{The models are atmosphere-only.} Sea surface temperature is prescribed from observations, so the models respond to ENSO but cannot generate it. Extending \ace to coupled atmosphere--ocean prediction \cite[e.g.][]{duncan26}, where the daily atmospheric time step would substantially reduce coupling cost, is a natural next step and work is ongoing. Using a prior stochastic atmosphere model (ACE2S) as a source of stochasticity for the ocean has already proved beneficial for SST variability in eddy-rich regions and for the spectrum of ENSO variability \citep{wu26}.
\item \emph{Training stops before the validation loss is minimized}, because longer pre-training degraded the multi-decadal climate of some seeds. This limits weather skill.
\end{itemize}

With the changes in \ace, degraded climate when training on the relatively small ERA5 dataset has been significantly reduced, but climate biases do still arise with sufficient training. We will continue to iterate on architectural design and training strategies to further align the optimization target with long-term climate skill.

We release three checkpoints that serve different applications. \ace (1-degree daily) is the primary model, with the best balance of climate skill and throughput. \ace-4deg (4-degree daily) has comparable climate skill at its target resolution, has 5.2$\times$ the throughput of \ace and trains in an order of magnitude fewer GPU hours, making it suitable for rapid development iterations and large-ensemble studies at reduced resolution. \ace-6h (1-degree 6-hourly) is intended for applications that require a diurnal cycle or competitive short-range weather skill.

\section*{Data and code availability}

The model checkpoints are available at \url{https://huggingface.co/allenai/ACE2.5-ERA5} under the Apache 2.0 license. The ERA5 training data (regridded to Gaussian grids) is at \url{https://huggingface.co/buckets/allenai/ACE2.5-ERA5-reference-data}. The figure reproduction code, evaluation configurations, and training configurations are archived at \url{https://doi.org/10.5281/zenodo.24406261} and developed at \url{https://github.com/ai2cm/ace2.5-era5-paper}. The ACE model library (fme) is at \url{https://github.com/ai2cm/ace}.

\section*{Acknowledgements}

This work was supported by the Fund for Science and Technology. ERA5 data were generated using Copernicus Climate Change Service information \citep{hersbach20}; neither the European Commission nor ECMWF is responsible for any use that may be made of the data. WeatherBench2 scorecards \citep{rasp24} were used to evaluate the weather forecasts of GraphCast and IFS HRES. Computations were performed on Ai2's Beaker platform using NVIDIA H100 GPUs. Thank you to Andre Perkins for managing shared cloud resources used for this project, to Boris Bonev for insightful discussions about the SFNO architecture, and to Anna Kwa and Elynn Wu for providing helpful reviews of code used in this project.

\appendix

\section{Appendix}

\subsection{Loss function details}
\label{eq:crps-definition}

CRPS in the negative orientation (where lower is better) \citep{gneiting07} is defined as
\begin{align*}
\mathrm{CRPS}(F, x)
= \int_{-\infty}^{\infty} \bigl(F(y) - \mathbf{1}\{y \ge x\}\bigr)^2 \, dy
\end{align*}
where $F$ is the cumulative distribution function for values in $\mathbb{R}$, and is shown to have an equivalent form of
\begin{align*}
    \text{CRPS}(F, y) \;=\; \mathbb{E}_{X \sim F}\!\left[\,|X - y|\,\right]
\;-\; \tfrac{1}{2}\,\mathbb{E}_{X, X' \sim F}\!\left[\,|X - X'|\,\right],
\end{align*}
where $F$ is the random distribution function of values in $\mathbb{R}$ and $y \in \mathbb{R}$ is a true sample.

Energy score \citep{gneiting07} is given by
\begin{align*}
    \text{ES}(F, \vec{y}) \;=\; \mathbb{E}_{\vec{X} \sim F} \big[ \, \lVert \vec{X} - \vec{y} \rVert \,\big]
\;-\; \tfrac{1}{2} \, \mathbb{E}_{\vec{X}, \vec{X}' \sim F} \big[ \, \lVert \vec{X} - \vec{X}' \rVert \,\big]
\end{align*}
and is defined for $y \in \mathbb{R}^n$. We apply the energy score independently to each complex coefficient of the spectral transform of the data, treated as vectors in $\mathbb{R}^2$, and then average it across these spectral coefficients. Note the norm of a complex value treated as a vector in $\mathbb{R}^2$ is identical to the complex modulus. In our case of two ensemble members, this can be written as

\begin{align}
  \overline{\text{ES}(\text{SHT} \circ F, \text{SHT}(y))}
  = \frac{1}{N_\ell N_m} \sum_{\ell, m} w_m \left[
    \tfrac{1}{2}\textstyle\sum_{i=1}^2\big|\hat{X}_{i,\ell m} - \hat{y}_{\ell m}\big|
    - \tfrac{1}{2}\big|\hat{X}_{1,\ell m} - \hat{X}_{2,\ell m}\big|
  \right],
  \label{eq:energy-score}
\end{align}

where $\hat{X}_{i} = \mathrm{SHT}(X_i)$, $|\cdot|$ is the complex modulus,
$N_\ell$ and $N_m$ are the number of retained degrees $\ell$ and orders $m$, and
$w_m = 1$ for $m = 0$ and $w_m = 2$ otherwise (the real SHT stores only
$m \geq 0$; doubling the $m > 0$ modes accounts for their conjugate
counterparts). The $1/(2\sqrt{N_\ell N_m})$ factor brings the term's magnitude
in line with the CRPS term. Because this term scores the distribution of the complex spectral coefficients themselves, evaluating both their real and imaginary components, it directly penalizes ensembles for discrepancies in both spectral power and phase at any wavenumber. This can be regarded as a `spectral CRPS'.

Our loss function is given by

\begin{align*}
    \mathcal{L}(F, y) = 0.9 \cdot \overline{\text{fCRPS}_{2}(F, y)} + 0.1 \cdot \frac{\sqrt{N_\ell N_m}}{2} \overline{\text{ES}(\text{SHT} \circ F, \text{SHT}(y))}
\end{align*}

where SHT is the spherical harmonic transform, with its complex-valued outputs treated as vectors in $\mathbb{R}^2$ for the purpose of computing the energy score, and expected values are calculated with an ensemble size of 2. Our decision to take the average value of CRPS without area-weighting best matches the choice of MSE loss without area-weighting for ACE2.

We use fair CRPS rather than the almost-fair variant of \citet{lang24}: the spectral loss term constrains the target even in the case of a perfectly accurate prediction at one point, so the ensemble collapse that almost-fair CRPS guards against did not arise.

\subsection{ERA5 dataset preparation}
\label{sec:si-data}

The ERA5 datasets used to train and evaluate \ace are new regriddings of ERA5 from the 0.25$^\circ$ native resolution, produced by a Dataflow pipeline that reads the ARCO-ERA5 archive \citep{arcoera5}. The pipeline regrids all fields to a Gaussian grid using area-conservative regridding (xESMF, \citet{xesmf}), producing a 6-hourly zarr store at each target resolution (1$^\circ$ and 4$^\circ$). The 6-hourly data are then time-coarsened to daily: analysis fields retain the 00Z snapshot every 24 hours, and window fields (fluxes, precipitation rate) are the mean of the four 6-hourly values.

The \ace datasets differ from the ACE2-ERA5 dataset of \citet{wattmeyer25} in three respects. First, the regridding method changed from MetView \citep{metview} to xESMF conservative regridding. Conservative regridding computes area-weighted averages of source cells, producing a smoother field at the target resolution than MetView's interpolation. This affects the grid-scale spectral power of the regridded fields: the \ace dataset retains less spectral power at small scales. For precipitation rate, the ACE2 dataset has approximately 50\% more spectral power at the highest resolved total wavenumber than the \ace dataset for the same years and resolution (Figure~\ref{fig:power_spectrum_comparison}). This is why each model in this paper is evaluated against the ERA5 dataset it was trained on, and why the ERA5 targets are shown separately in figures that compare precipitation at grid scale. Second, the \ace datasets extend through 2025, adding three years of training data. Third, the \ace datasets include additional output variables needed by the expanded model (mean sea level pressure, surface wind stresses, frozen precipitation rate, and seven daily-mean near-surface diagnostics). The dataset also includes clear-sky radiative fluxes, soil temperatures, snow fields and surface wave height, though these are unused in this work.

The 4-degree datasets are produced by the same pipeline at F22.5 Gaussian resolution (45 latitudes, 90 longitudes), with daily time-coarsening applied identically. No 4-degree ERA5 dataset existed for ACE2.

\subsection{Finite-ensemble RMSE adjustment}
\label{sec:si-ensemble-stats}

For a calibrated ensemble whose members are exchangeable with the truth, the expected squared error of a single member is twice the forecast variance $\sigma^2$, while that of the $M$-member mean is $(1 + 1/M)\,\sigma^2$. The ratio of single-member to ensemble-mean RMSE is therefore $\sqrt{2M/(M+1)}$, which evaluates to 1.34 for $M = 9$. The measured ratio at day~5 is between 1.33 and 1.35 in every field for \ace (1.32--1.33 for \ace-4deg and 1.34--1.38 for \ace-6h), consistent with calibrated ensembles. By the same argument the $M$-member mean RMSE overstates the infinite-ensemble ($M \to \infty$) value by a factor $\sqrt{(M+1)/M} \approx 1.05$ for $M = 9$. The ``skill'' plotted in the calibration figure (Figure~\ref{fig:calibration}, top) is this corrected quantity: the RMSE of the ensemble mean divided by $\sqrt{(M+1)/M}$.

\subsection{Evaluation details}
\label{sec:si-diagnostics}

\begin{figure}
\centering
\includegraphics[width=\textwidth]{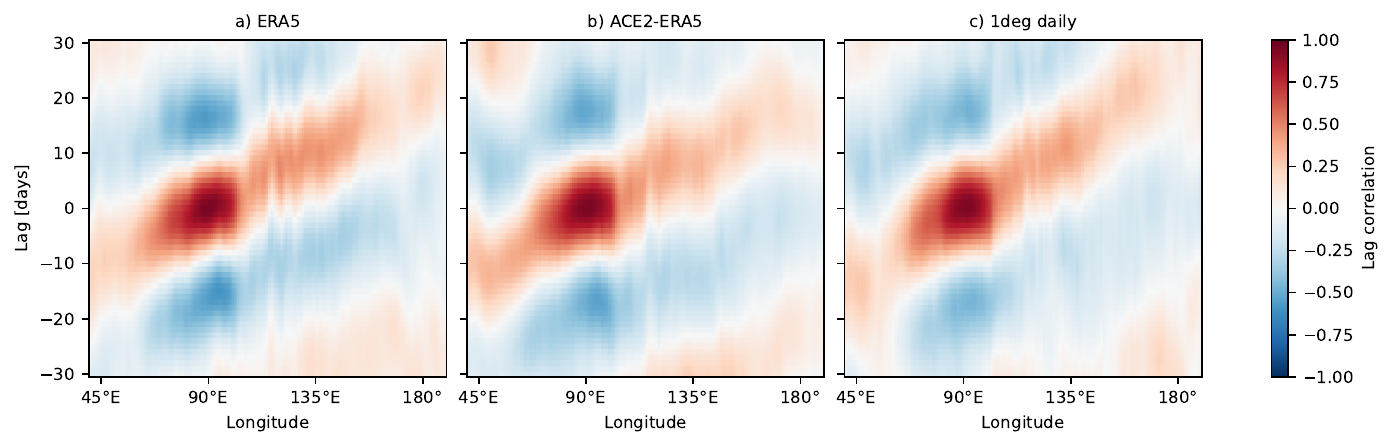}
\caption{Lag correlation of 20--100-day filtered tropical precipitation against its 80--100$^\circ$E mean, over the 10-year runs, for ERA5, \ace and ACE2-ERA5.}
\label{fig:mjo}
\end{figure}

\begin{figure}
\centering
\includegraphics[width=\textwidth]{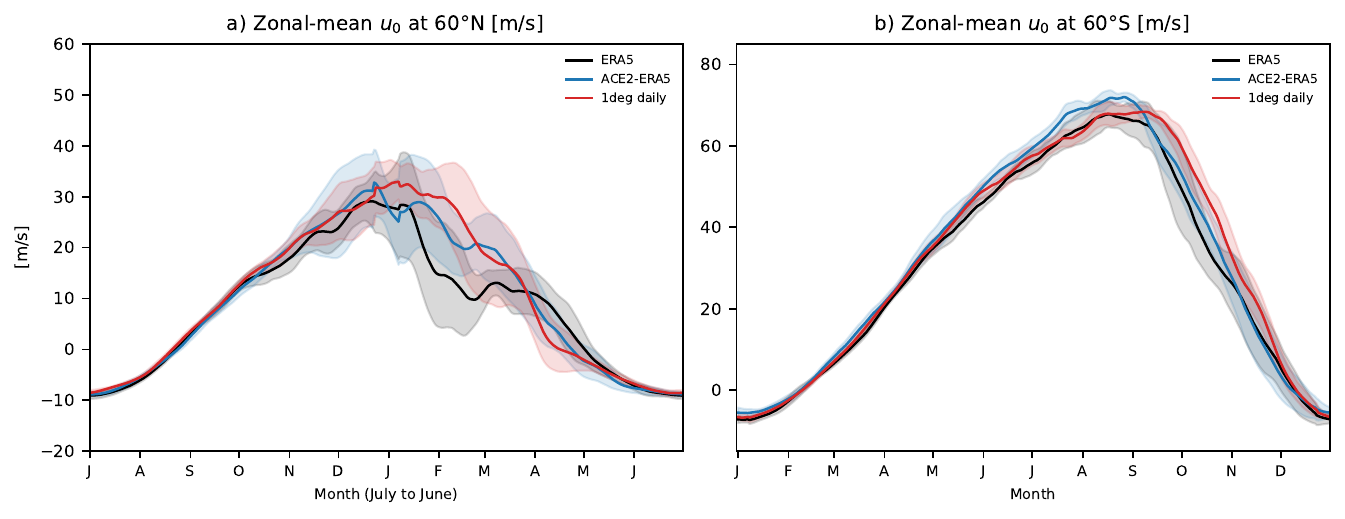}
\caption{Zonal-mean zonal wind at the top model level at 60$^\circ$N (a, July to June) and 60$^\circ$S (b) over the 10-year runs. Solid lines are the day-of-year mean after 15-day smoothing; shading is the 95\% confidence interval of the mean across the 10 years.}
\label{fig:strat}
\end{figure}

Figures~\ref{fig:hovmoller}, \ref{fig:mjo} and \ref{fig:strat} repeat the checks of tropical variability and of the stratospheric annual cycle from \citet{wattmeyer25} for the first seed of \ace, using its 10-year run initialized on 2001-01-01. The ACE2-ERA5 fields are its 6-hourly 10-year run averaged to daily means, and ERA5 is the daily-mean precipitation of the training dataset over the same years.

\paragraph{Madden--Julian Oscillation.} For Figure~\ref{fig:mjo}, the precipitation averaged over 10$^\circ$S--10$^\circ$N (shown for 2008 in Figure~\ref{fig:hovmoller}) is band-pass filtered to 20--100-day periods (a fifth-order Butterworth filter) over the full 10 years and correlated, at lags of up to 30 days, against its mean over 80--100$^\circ$E. \ace reproduces the eastward propagation of the Madden--Julian Oscillation across the Indian Ocean and the Maritime Continent, with a weaker signal east of 120$^\circ$E than either ERA5 or ACE2-ERA5.

\paragraph{Stratospheric vortex.} Figure~\ref{fig:strat} shows the zonal-mean zonal wind of the topmost model layer at 60$^\circ$N and 60$^\circ$S, with 95\% confidence intervals of the 10-year day-of-year mean (after 15-day smoothing); here ACE2-ERA5 is the 10-year run that saved daily output, and ERA5 the 00Z values of the training dataset. \ace and ACE2-ERA5 both reproduce the annual cycle of the Antarctic vortex, with confidence intervals that overlap ERA5's throughout the year. In the Arctic, both tend to hold the vortex stronger than ERA5 through February--March, but the difference is not statistically significant: sudden stratospheric warmings make the 10-year mean uncertain, and the few points of near-separation are consistent with the number of false positives expected over the annual cycle.

\begin{figure}
\centering
\includegraphics[width=0.75\textwidth]{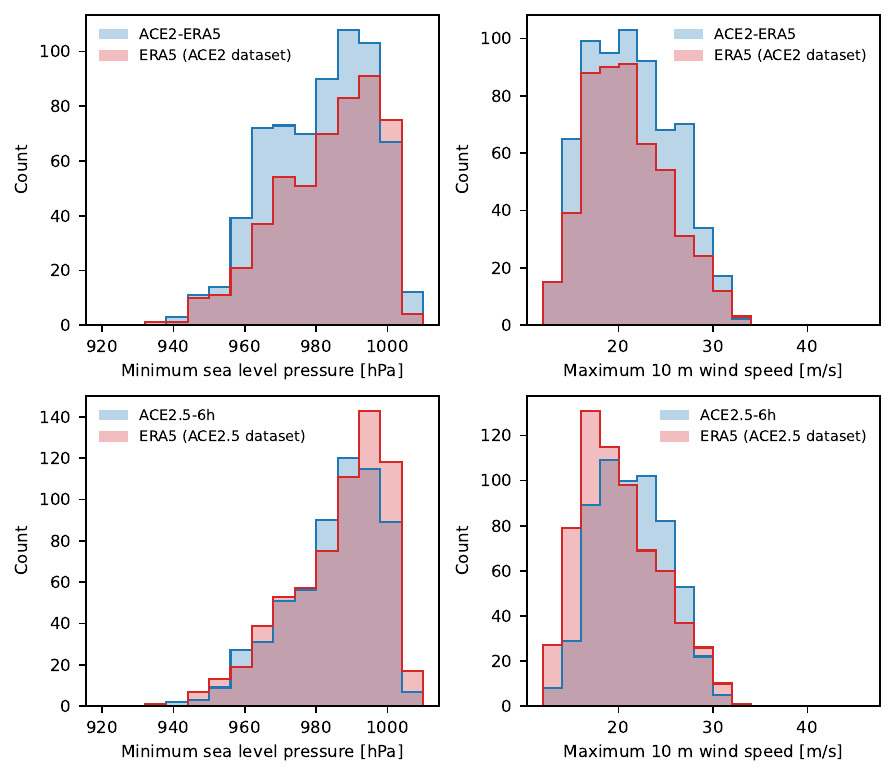}
\caption{Distribution of tropical cyclone intensity over the 10-year evaluation period: lifetime minimum sea level pressure (left) and lifetime maximum 10\,m wind speed (right) for each detected storm. Top: ACE2-ERA5 against the ACE2 ERA5 dataset; bottom: \ace-6h against the \ace ERA5 dataset.}
\label{fig:tc_intensity}
\end{figure}

\paragraph{Tropical cyclone tracking.} The tracks of Figures~\ref{fig:tc_tracks} and \ref{fig:tc_intensity} are identified with TempestExtremes \citep{ullrich21}. A candidate is a sea level pressure minimum with a 200\,Pa closed contour within 5.5$^\circ$ and a warm core (temperature at model level 3 decreasing by 0.4\,K over 6.5$^\circ$); candidates within 6$^\circ$ are merged. Candidates are stitched into tracks with a maximum step of 8$^\circ$ and a maximum gap of 24\,h. A track must last at least 54\,h and, for at least 10 of its 6-hourly points, have a 10\,m wind speed of at least 10\,m/s, a latitude within 50$^\circ$ of the equator and a surface height below 150\,m.

\subsection{Ablation studies}
\label{sec:ablation-figures}

Table~\ref{tab:ablations} lists every ablation model named in this paper. Each is derived from \ace or \ace-6h by the change listed; the section cited gives its full configuration, including any difference in training length.

\begin{table}
\centering
\caption{Ablation models, the model each is derived from, and the change that defines it.}
\label{tab:ablations}
\small
\begin{tabular}{lllp{0.4\textwidth}}
\toprule
Model & Base & Section & Change \\
\midrule
\ace-deterministic & \ace & \ref{sec:si-loss-ablation} & No noise input; MSE loss with ACE2-ERA5's per-variable weights; one deterministic prediction \\
\ace-crps-only & \ace & \ref{sec:loss} & CRPS loss alone, without the spectral energy score \\
\ace-no-bottleneck & \ace & \ref{sec:bottlenecking} & \texttt{spectral\_ratio} of 1 and a single group \\
\ace-6h-svfs & \ace-6h & \ref{sec:training} & Fine-tuned on a stochastically varying number of forward steps \citep{perkins26} \\
\ace-embed384 & \ace & \ref{sec:si-weather-gap} & Embedding dimension of 384 \\
\ace-embed1024 & \ace & \ref{sec:si-weather-gap} & Embedding dimension of 1024 \\
\ace-6h-samples & \ace & \ref{sec:si-weather-gap} & Daily step, trained on all four 6-hourly phases \\
\ace-6h-quarter & \ace-6h & \ref{sec:si-weather-gap} & Trained on a quarter of the years \\
\bottomrule
\end{tabular}
\end{table}

\subsubsection{Loss function, bottleneck and fine-tuning schedule}
\label{sec:si-loss-ablation}

\begin{figure}
\centering
\includegraphics[width=0.8\textwidth]{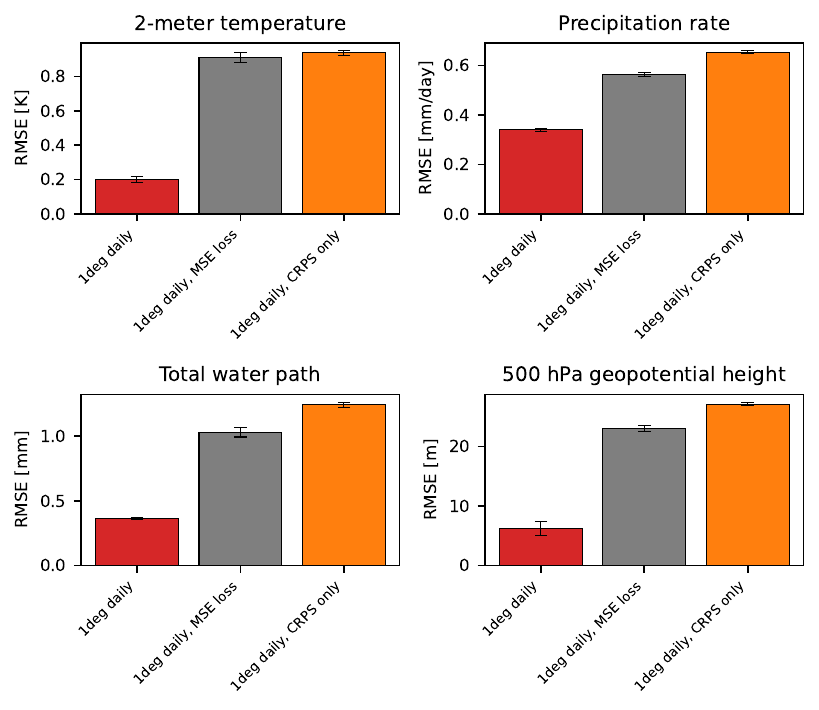}
\caption{Area-weighted RMSE of the 10-year time mean against ERA5 for \ace, \ace-deterministic (MSE loss) and \ace-crps-only (CRPS loss only).}
\label{fig:si_loss_rmse}
\end{figure}

\begin{figure}
\centering
\includegraphics[width=0.7\textwidth]{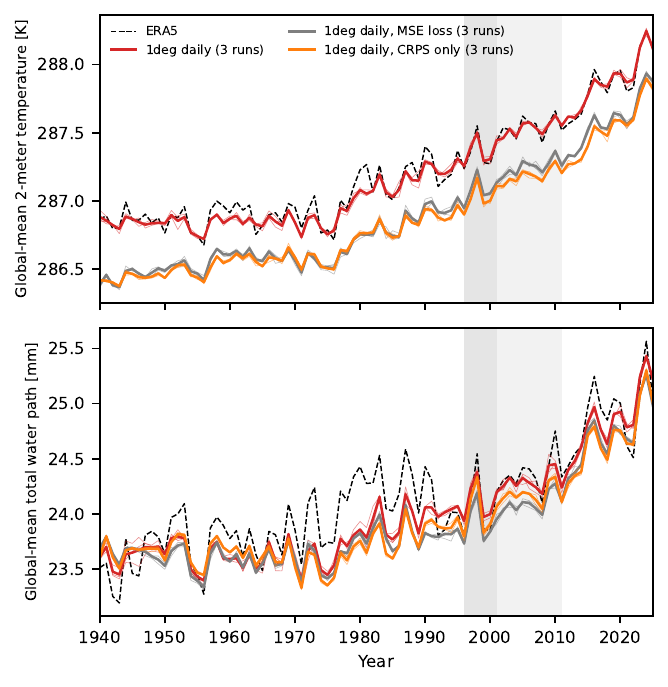}
\caption{Annual global-mean 2\,m temperature (top) and total water path (bottom) over the AMIP runs for \ace, \ace-deterministic and \ace-crps-only, compared with ERA5 (dashed).}
\label{fig:si_loss_climate}
\end{figure}

\begin{figure}
\centering
\includegraphics[width=\textwidth]{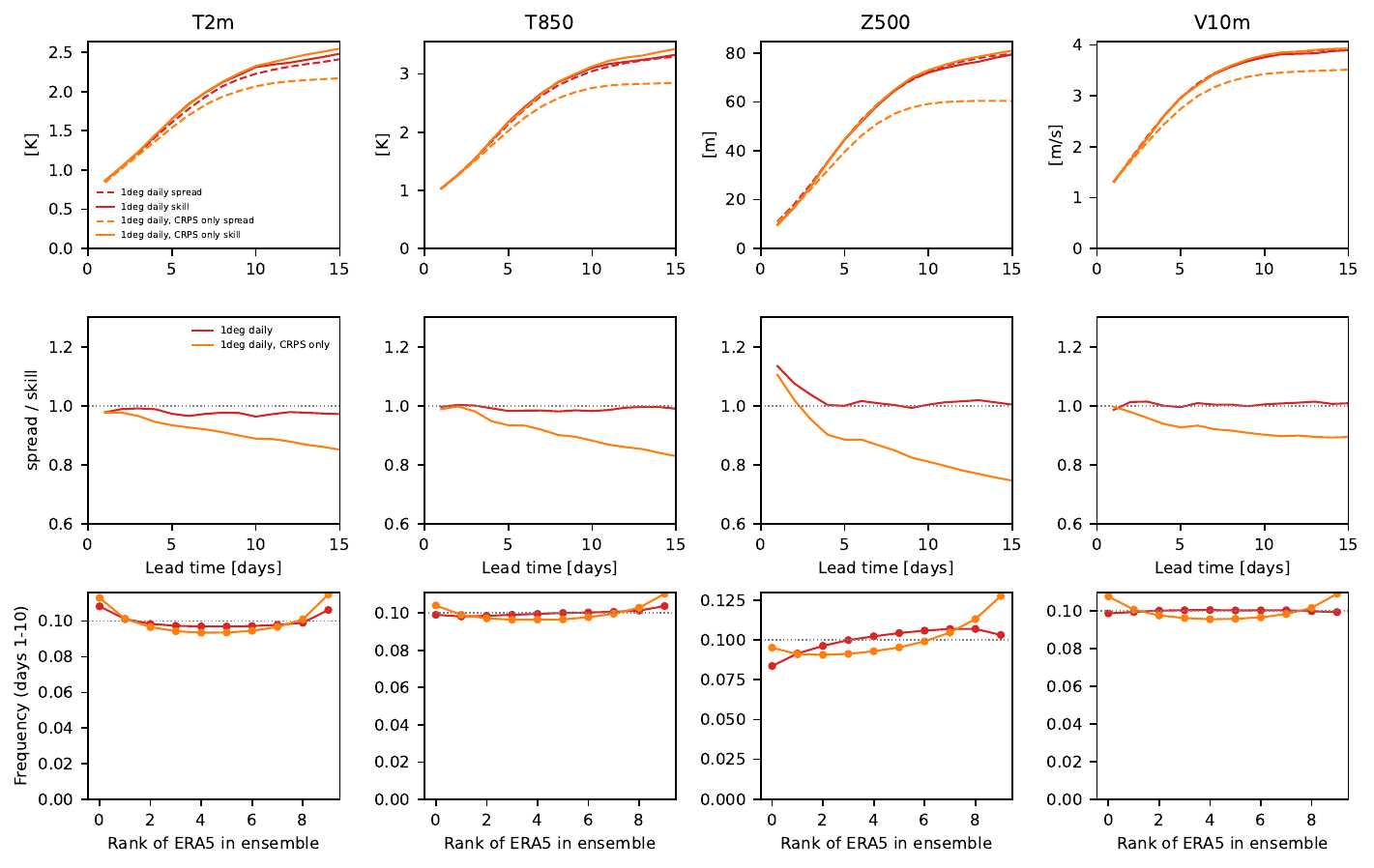}
\caption{Ensemble calibration of \ace and \ace-crps-only: spread and skill (top), their ratio (middle) and rank histogram pooled over days 1--10 (bottom). The deterministic model is excluded as it produces a single member.}
\label{fig:si_loss_calibration}
\end{figure}

\begin{figure}
\centering
\includegraphics[width=\textwidth]{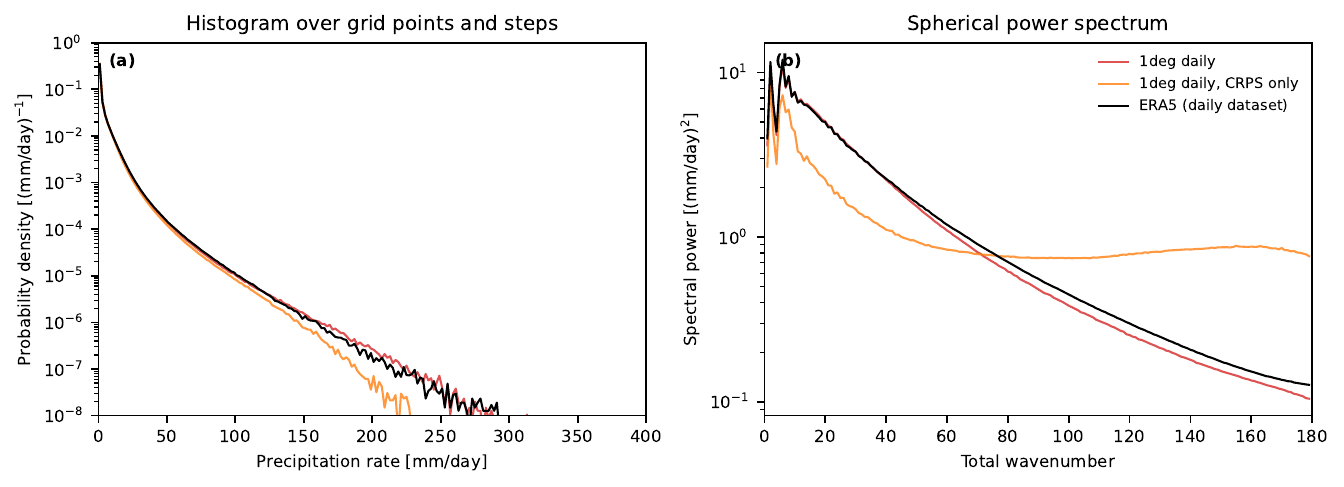}
\caption{Precipitation histogram (a) and spherical power spectrum (b) for \ace and \ace-crps-only, compared with ERA5 (dashed).}
\label{fig:si_loss_precip}
\end{figure}

We compare \ace (CRPS + spectral energy score) with two loss function ablations, \ace-deterministic and \ace-crps-only. Both degrade climate fidelity. \ace-deterministic produces a smoother, less realistic precipitation field and a larger time-mean temperature RMSE (Figure~\ref{fig:si_loss_rmse}). \ace-crps-only develops excess spectral power at small scales (Figure~\ref{fig:si_loss_precip}), as shown in Section~\ref{sec:loss}, and drifts in multi-decadal climate runs (Figure~\ref{fig:si_loss_climate}). CRPS-only training also produces an under-dispersed ensemble (Figure~\ref{fig:si_loss_calibration}), with the spread-to-skill ratio 0.89--0.93 at day 5 and 0.75--0.89 by day 15.

\begin{figure}
\centering
\includegraphics[width=0.85\textwidth]{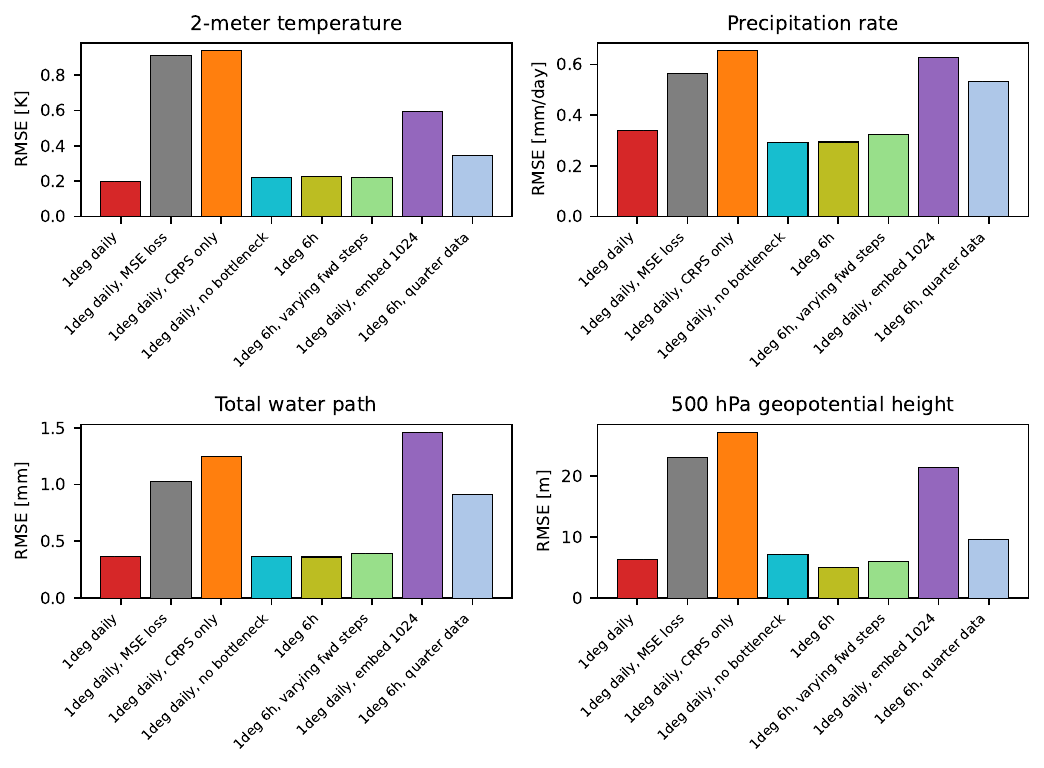}
\caption{Area-weighted RMSE of the 10-year time mean against ERA5 for all ablation models, including \ace-embed1024 and \ace-6h-quarter from Appendix~\ref{sec:si-weather-gap}. Daily-step models are scored on daily means, 6-hourly models on the mean of their 6-hourly snapshots. }
\label{fig:si_ablation_rmse}
\end{figure}

\begin{figure}
\centering
\includegraphics[width=\textwidth]{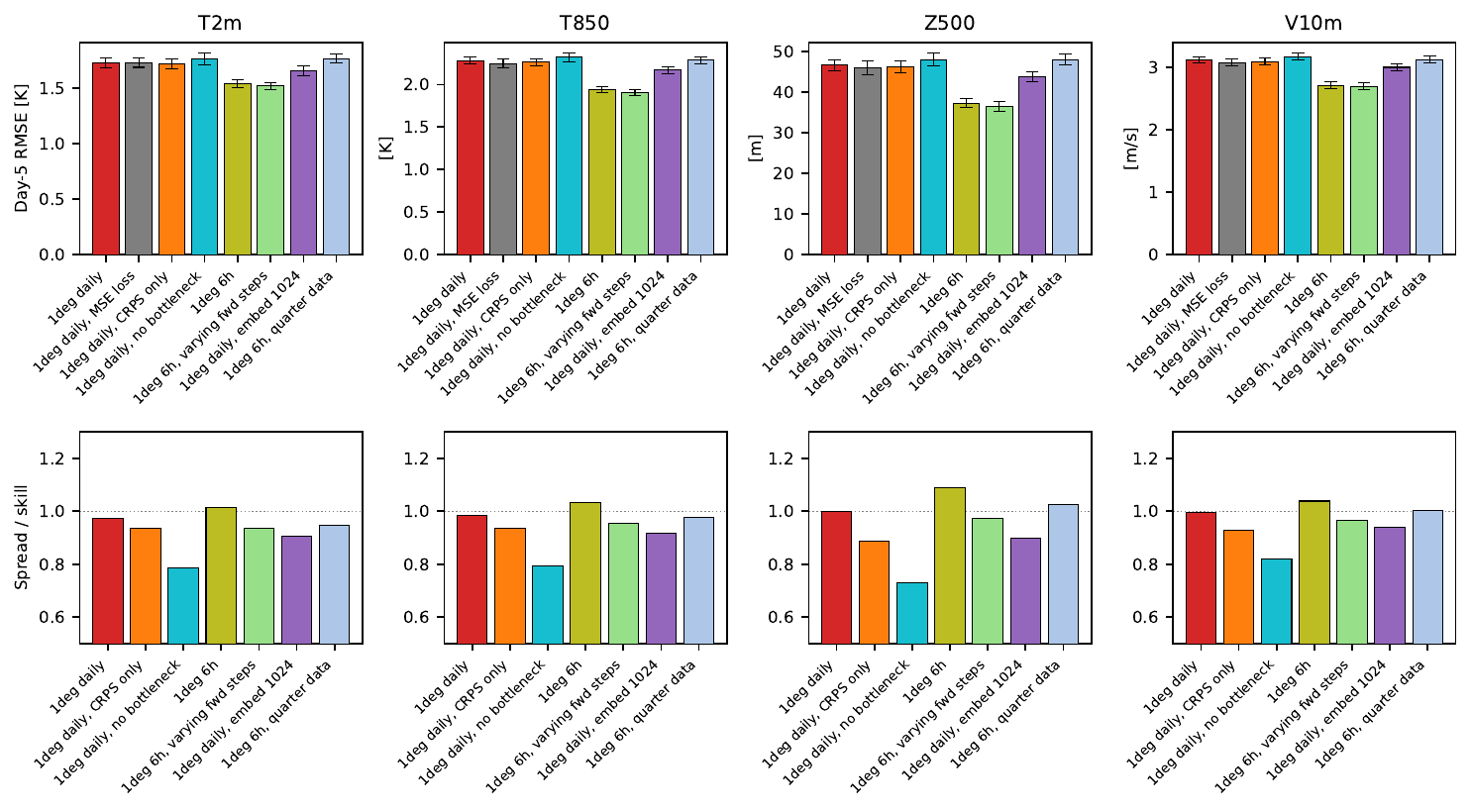}
\caption{Day-5 ensemble-mean RMSE (top; single-member RMSE for the deterministic model) and spread-to-skill ratio (bottom; stochastic models only) for all ablation models. A ratio of 1 is perfectly calibrated.}
\label{fig:si_ablation_weather}
\end{figure}

The other ablations leave climate fidelity largely intact. \ace-no-bottleneck matches \ace in time-mean RMSE and weather skill (Figures~\ref{fig:si_ablation_rmse} and~\ref{fig:si_ablation_weather}) but is under-dispersed at short leads (Figure~\ref{fig:bottleneck_calibration}). \ace-6h-svfs matches \ace-6h in weather skill (Figure~\ref{fig:si_ablation_weather}) with similar climate fidelity (Figure~\ref{fig:si_ablation_rmse}). Two of the configurations trained to explain the daily model's weather skill gap (Appendix~\ref{sec:si-weather-gap}) have worse climate fidelity. \ace-embed1024 has better weather skill than \ace, but its time-mean RMSE is far larger in every field (0.59 against 0.20\,K in 2\,m temperature, 1.46 against 0.36\,mm in total water path). \ace-6h-quarter is worse than \ace-6h in both weather skill and time-mean RMSE (0.35 against 0.22\,K).

\paragraph{Spectral ratio sensitivity.} We also trained models at 1-degree daily resolution with \texttt{spectral\_ratio} values of 1/2, 1/4 and 1/16, keeping all other hyperparameters identical to \ace. The 10-year time-mean 2\,m temperature RMSE is 0.24\,K for the paper model (1/8) and within 0.02\,K of that for every other ratio. Total water path and 500\,hPa height have slightly better time-mean RMSEs at wider bottlenecks, but the differences are within the 95\% confidence interval across the three initial conditions. Weather forecast RMSE varies by less than 3\% across the four ratios in every field. In 81-year AMIP runs all four ratios track the observed warming without drift; the paper model has the lowest 2\,m temperature RMSE (0.109\,K) and the narrowest ratio (1/16) the highest (0.137\,K). The 1/8 ratio thus minimizes parameter count (46M vs.\ 59M at 1/2) with no skill penalty, confirming that the value tuned at 4-degree resolution transfers to 1 degree.

\subsubsection{Weather skill gap}
\label{sec:si-weather-gap}

Section~\ref{sec:weather} notes that the daily timestep model has worse weather skill relative to \ace-6h, and that the 6-hourly model uses four times as much compute for the same number of training epochs. We test three hypotheses for the gap: (i) it reflects a deficit in optimizer steps, since the daily model's 40-epoch pre-training amounts to 128k gradient updates versus 511k for the 6-hourly model; (ii) the daily model has fewer training samples per epoch, seeing only the 00Z snapshot while the 6-hourly model sees all four 6-hourly phases; (iii) the model's width limits skill at a given step budget.

\begin{figure}
\centering
\includegraphics[width=\textwidth]{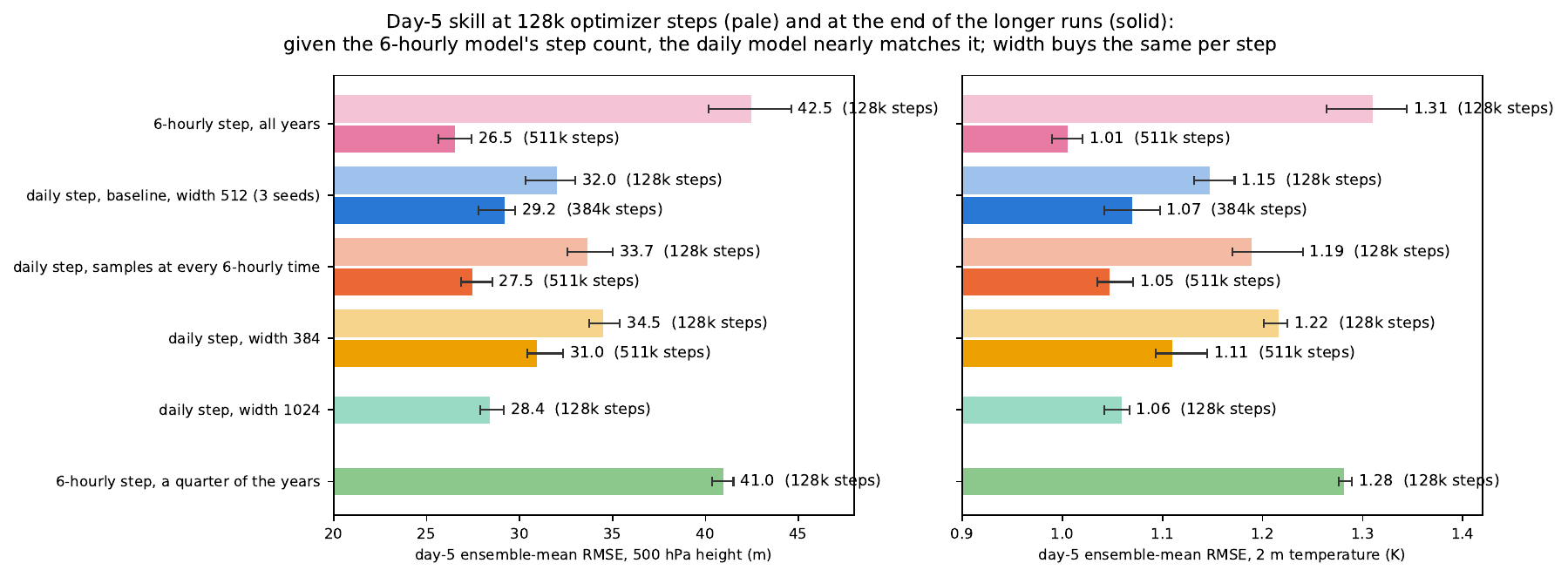}
\caption{Day-5 ensemble-mean RMSE from the inline 5-year evaluation during pre-training (8 initial conditions in 1996, 3 ensemble members), at 128k optimizer steps (pale bars) and at the end of the longer runs (solid bars). Error bars show the range over the five training epochs nearest the target step count. At matched step counts, the daily recipe's 500\,hPa height skill is 25\% better than the 6-hourly model's; wider embedding (1024) improves skill per step; training on all four 6-hourly phases does not help at equal steps. The daily-baseline bars average pre-training seeds 1--3: the seed-0 pre-training run, from which the released \ace is fine-tuned, predates the inline ensemble-mean diagnostic.}
\label{fig:weather_gap_verdict}
\end{figure}

Figure~\ref{fig:weather_gap_verdict} shows the day-5 RMSE of each ablation during pre-training, evaluated at 128k optimizer steps (the daily recipe's 40 epochs) and at the end of longer runs. At 128k steps, the daily model's 500\,hPa height RMSE (32\,m) is 25\% better than the 6-hourly model at the same step count (42\,m). Training the daily model on all four 6-hourly phases (\ace-6h-samples) raises the per-epoch step count to match the 6-hourly model's, but does not help at equal step counts (34\,m at 128k steps). Training a 6-hourly model on a quarter of the years (matching the daily model's step count per epoch) gives similar skill to the full 6-hourly model at equal steps (41\,m at 128k). These results isolate the gap as a step deficit, not a sample-count or temporal-resolution effect.

\begin{figure}
\centering
\includegraphics[width=\textwidth]{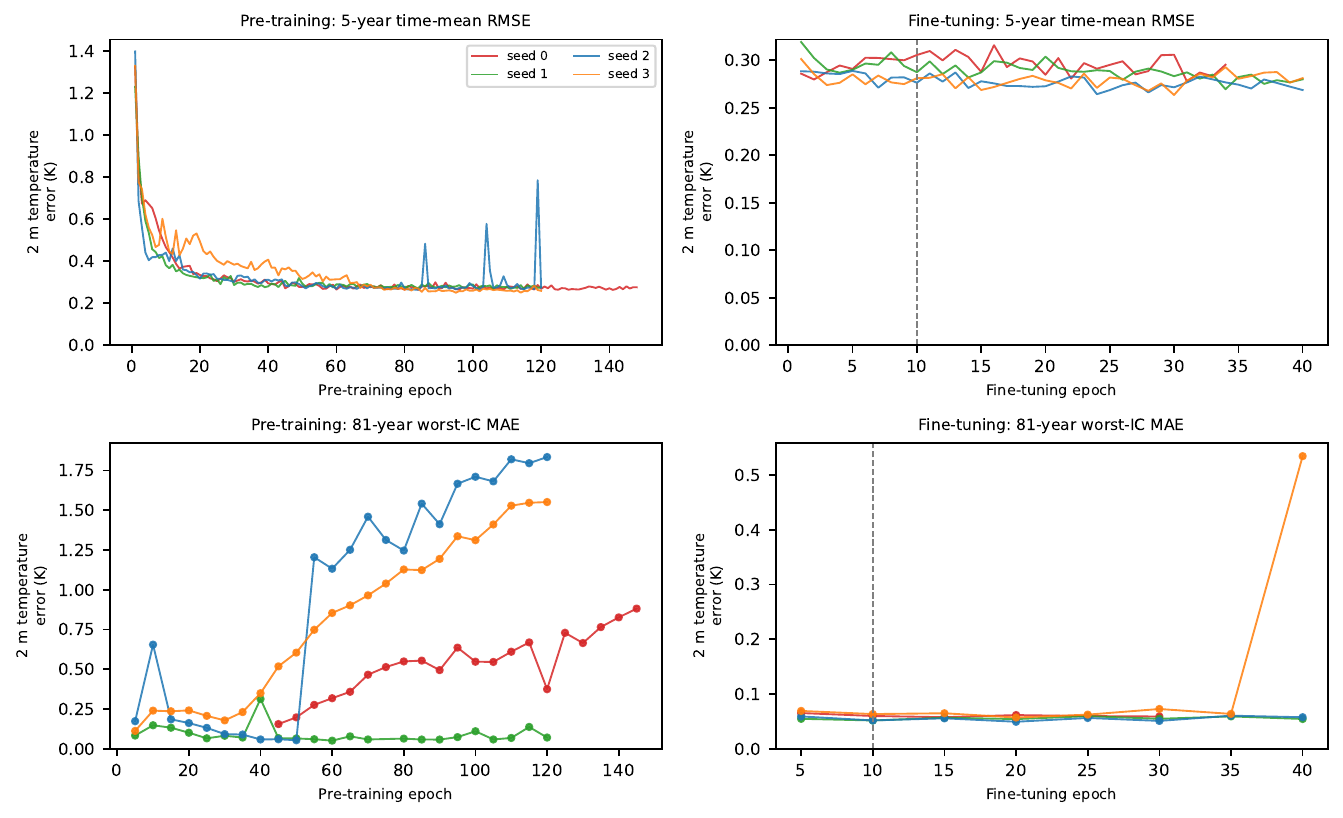}
\caption{Climate drift during extended training for all four random seeds, evaluated inline from eight initial conditions. Top row: area-weighted RMSE of the 5-year time-mean 2\,m temperature (scored every epoch). Bottom row: worst initial condition's mean absolute error of annual-mean 2\,m temperature over an 81-year rollout (scored every five epochs). The 5-year RMSE (top) plateaus by epoch 20 for all seeds, while the 81-year worst-IC error (bottom left) grows with continued pre-training, with two seeds exceeding 1.5\,K by epoch 120. The onset and severity differ by seed. Fine-tuning (right) is stable through 30 epochs for all seeds. The dashed line marks the selected fine-tuning epoch (10).}
\label{fig:training_drift}
\end{figure}

Increasing the embedding dimension from 512 to 1024 improves skill within a fixed step budget: the 1024-wide daily model at 128k steps (28\,m) nearly matches the 512-wide daily model trained for three times as long (29\,m at 384k steps). It is possible that a 512-wide daily model with both good weather and climate skill could be trained, however we found that for some epochs and seeds later in training, the AMIP inference runs had significant climate biases in some initializations, and those biases were not present in the shorter 5-year inline inference runs (Figure~\ref{fig:training_drift}). This would require care in selecting seed and epoch. Because of the focus of this work on climate skill and on comparison of perturbed configurations, we stop pre-training at 40 epochs.

The day-5 RMSE and spread-to-skill ratios for these and all other ablation models are in Figure~\ref{fig:si_ablation_weather}. All stochastic ablation models produce well-calibrated ensembles at day 5.

\subsection{Random seed variability}
\label{sec:seed-variability}

\ace and \ace-4deg are each trained with four random seeds, and each seed is evaluated with the full protocol of Table~\ref{tab:protocol}. All figures in the main text show seed 0; here we show that the other seeds produce comparable results.

\begin{figure}
\centering
\includegraphics[width=0.85\textwidth]{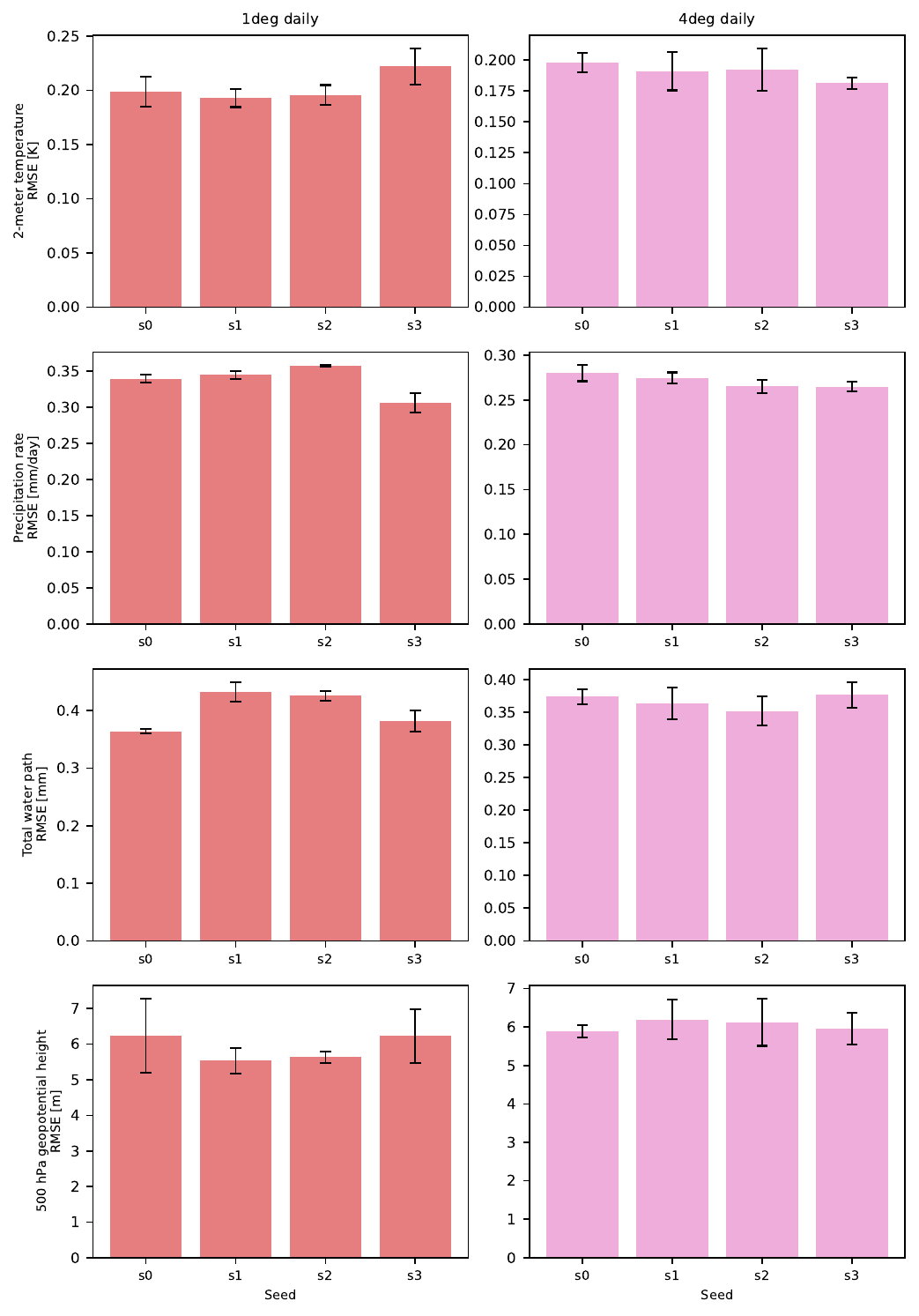}
\caption{Time-mean RMSE by random seed for \ace (left) and \ace-4deg (right), across the four climate variables of Figure~\ref{fig:time_mean_rmse}. Each bar is the RMSE averaged over the three initial conditions.}
\label{fig:seed_rmse}
\end{figure}

\begin{figure}
\centering
\includegraphics[width=\textwidth]{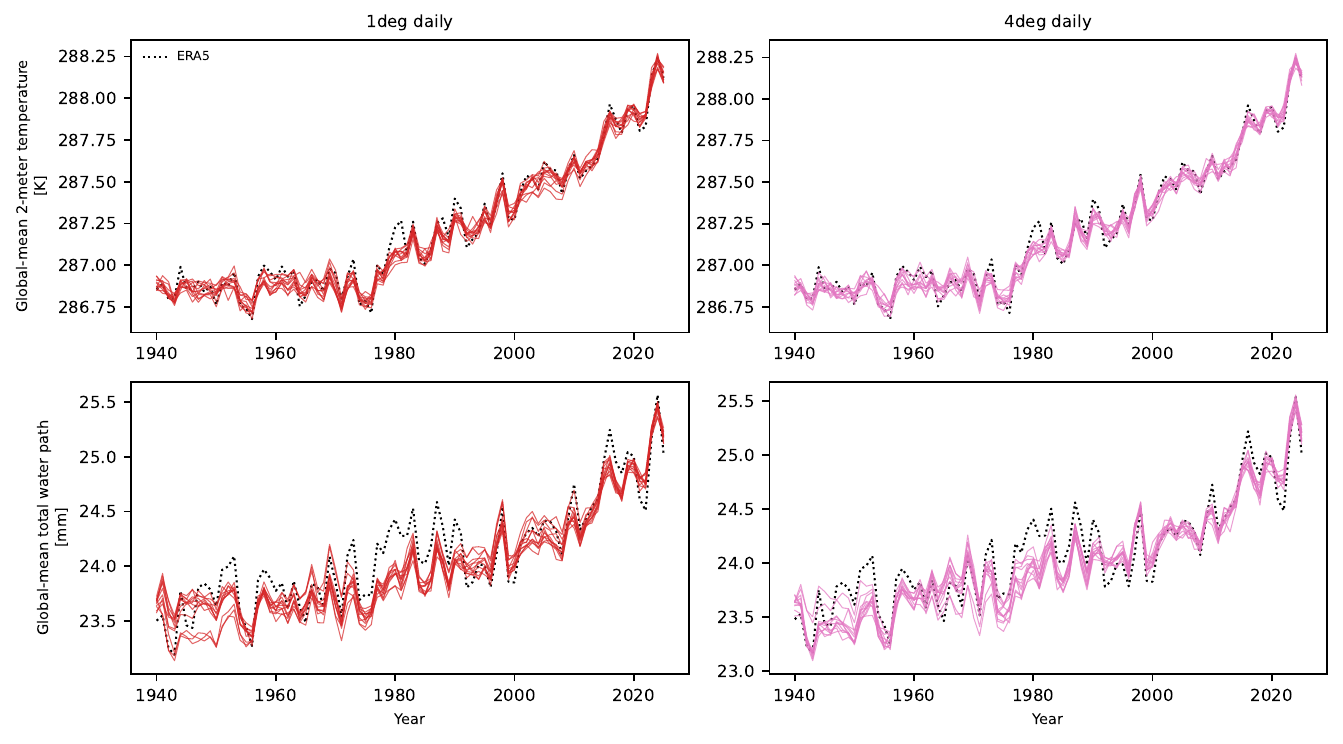}
\caption{AMIP climate evolution by random seed for \ace (left) and \ace-4deg (right). Thin lines show individual seed $\times$ initial condition runs; ERA5 is dotted.}
\label{fig:seed_evolution}
\end{figure}

\begin{figure}
\centering
\includegraphics[width=\textwidth]{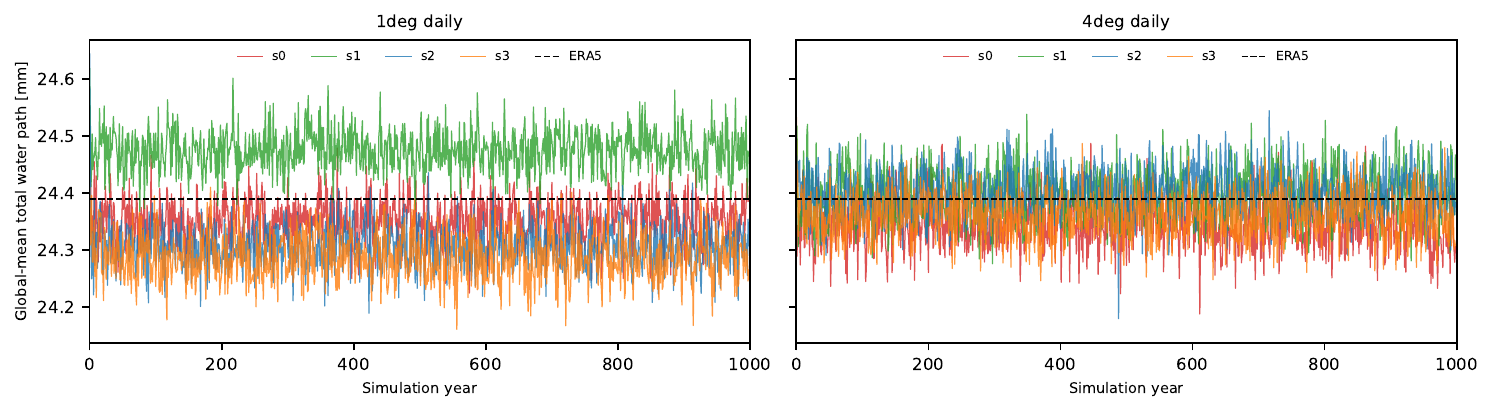}
\caption{1000-year total water path by random seed for \ace (left) and \ace-4deg (right). All four seeds of \ace and of \ace-4deg stay within 0.11\,mm of ERA5's 1991--2020 mean (dashed).}
\label{fig:seed_1000yr}
\end{figure}

The time-mean RMSE varies by less than 20\% across seeds for every field, the worst case being \ace's total water path (Figure~\ref{fig:seed_rmse}). The AMIP climate evolution is visually indistinguishable across seeds (Figure~\ref{fig:seed_evolution}), and all eight 1000-year runs are stable with no drift (Figure~\ref{fig:seed_1000yr}).
\clearpage
\bibliographystyle{plainnat}
\bibliography{references}

\end{document}